\documentclass[12pt]{article}
\usepackage{amsmath}
\usepackage{graphicx}
\usepackage{enumerate}
\usepackage{natbib}
\usepackage{url} 

\usepackage{geometry}
\usepackage{pdfpages}

\usepackage{amsthm}
\usepackage{amssymb}
\usepackage{xargs}[2008/03/08]
\usepackage{xcolor}
\usepackage{enumitem}
\usepackage{bm}
\usepackage{natbib}
\usepackage[plain,noend]{algorithm2e}
\usepackage{authblk}
\usepackage{float}

\theoremstyle{plain}

\newtheorem{theorem}{Theorem}[section]

\newtheorem{proposition}[theorem]{Proposition}

\theoremstyle{remark}

\newtheorem{assumption}[theorem]{Assumption}
\newtheorem{rem}{Remark}

\makeatletter
\renewcommand{\algocf@captiontext}[2]{#1\algocf@typo. \AlCapFnt{}#2} 
\def\@algocf@capt@plain{top}
\renewcommand{\algocf@makecaption}[2]{%
	\addtolength{\hsize}{\algomargin}%
	\sbox\@tempboxa{\algocf@captiontext{#1}{#2}}%
	\ifdim\wd\@tempboxa >\hsize
	\hskip .5\algomargin%
	\parbox[t]{\hsize}{\algocf@captiontext{#1}{#2}}
	\else%
	\global\@minipagefalse%
	\hbox to\hsize{\box\@tempboxa}
	\fi%
	\addtolength{\hsize}{-\algomargin}%
}
\makeatother

\makeatletter
\newcommand{\ostar}{\mathbin{\mathpalette\make@circled\star}}
\newcommand{\make@circled}[2]{%
	\ooalign{$\m@th#1\smallbigcirc{#1}$\cr\hidewidth$\m@th#1#2$\hidewidth\cr}%
}
\newcommand{\smallbigcirc}[1]{%
	\vcenter{\hbox{\scalebox{0.77778}{$\m@th#1\bigcirc$}}}%
}
\makeatother

\def\expect{\mathbb{E}}
\def\prob{\mathbb{P}}
\def\diffop{\mathrm{d}}
\def\real{\mathbb{R}}

\def\transpose{\top}

\def\xinnerprod#1#2{\langle#1,#2\rangle_{1}}
\def\yinnerprod#1#2{\langle#1,#2\rangle_{2}}
\newcommand\xnorm[1]{\|#1\|_{1}}
\newcommand\ynorm[1]{\|#1\|_{2}}

\def\tdomain{\mathcal{T}}

\def\kronprod{\ostar}
\def\P{\mathcal{P}}
\def\Tn{h_n}

\usepackage[hidelinks]{hyperref}
\hypersetup{
	colorlinks,
	linkcolor={red!50!black},
	citecolor={blue!50!black},
	urlcolor={blue!80!black}
}

\usepackage{xr}
\makeatletter
\newcommand*{\addFileDependency}[1]{
	\typeout{(#1)}
	\@addtofilelist{#1}
	\IfFileExists{#1}{}{\typeout{No file #1.}}
}
\makeatother
\newcommand*{\myexternaldocument}[1]{%
	\externaldocument{#1}%
	\addFileDependency{#1.tex}%
	\addFileDependency{#1.aux}%
}
\myexternaldocument{iFLR-supp-arxiv}

\def\xref#1{\ref*{#1}}

\usepackage{microtype}

\date{}

\begin{document}

\def\spacingset#1{\renewcommand{\baselinestretch}%
	{#1}\small\normalsize} \spacingset{1}


\title{\bf Hypothesis Testing for Functional Linear Models via Bootstrapping}
\author{Yinan Lin\thanks{stayina@nus.edu.sg} }
\author{Zhenhua Lin\thanks{linz@nus.edu.sg}}
\affil{Department of Statistics and Data Science, National University of Singapore}
\maketitle

\begin{abstract} Hypothesis testing for the slope function in functional linear regression is of both practical and theoretical interest. We develop a novel test for the nullity of the slope function,  where testing the slope function is transformed into testing a high-dimensional vector based on functional principal component analysis. This transformation  fully circumvents ill-posedness in functional linear regression, thereby enhancing  numeric stability. The proposed method leverages the technique of bootstrapping max statistics and exploits the inherent variance decay property of functional data, improving the empirical power of tests especially when the sample size is limited or the signal is relatively weak. We establish validity and consistency of our proposed test when the functional principal components are derived from data. Moreover, we show that the  test maintains its asymptotic validity and consistency, even when including \emph{all} empirical functional principal components in our test statistics. This sharply contrasts with the task of estimating the slope function, which requires a delicate choice of the number (at most in the order of $\sqrt n$) of functional principal components to ensure estimation consistency. This distinction  highlights an interesting difference between estimation and statistical inference regarding the slope function in functional linear regression. To the best of our knowledge, the proposed test is the first of its kind to utilize all empirical functional principal components. 
\end{abstract}

\noindent%
\vfill

\newpage
\spacingset{1.3} 


\section{Introduction}\label{sec:intro}

Functional data  are nowadays common in practice and have been extensively studied in the past decades. For a comprehensive treatment on the subject of functional data analysis, we recommend the monographs \cite{Ramsay2005} and \cite{Kokoszka2017} for an introduction, \cite{Ferraty2006} for nonparametric functional data analysis, \cite{Hsing2015} from a theoretical perspective,  and  \cite{Horvath2012} and \cite{Zhang2013} with a focus on statistical inference.

Functional linear models that pair a response variable with a predictor variable in a linear way, where at least one of the variables is a function, play an important role in functional data analysis. A functional linear model (FLM), in its general form that accommodates both functional responses and/or functional predictors, can be mathematically represented by 
\begin{equation}
	Y - \expect Y = \beta(X - \expect X) + Z,
	\label{FLM}
\end{equation}
where $Y,Z\in\mathcal Y$, $X\in\mathcal X$, 
{with $(\mathcal{X}, \xinnerprod{\cdot}{\cdot})$ and $(\mathcal{Y}, \yinnerprod{\cdot}{\cdot})$ being two separable Hilbert spaces respectively endowed with the inner products $\xinnerprod{\cdot}{\cdot}$ and $\yinnerprod{\cdot}{\cdot}$,} 
and $\beta$, called the slope operator, is an unknown Hilbert--Schmidt operator between $\mathcal{X}$ and $\mathcal{Y}$. The variable $Z$, representing 
{a random error, }
is assumed to be centered, of finite variance, and independent of $X$. The model \eqref{FLM} includes the following popular models as special cases: the scalar-on-function model with a scalar response and a functional predictor, the function-on-function model with both functional response and predictor, the function-on-vector model (also known as the varying coefficient model in \cite{Shen2004}) with a functional response and multiple scalar predictors, the model with mixed-type predictors \citep{cao2020estimation}, and the partial functional linear model \citep{shin2009partial}; see Section \xref{sec:special-flm} of \cite{lin2024supphypothesis} for more details. These models have been investigated, for example, among many others, by   \cite{Cardot1999,Cardot2003,Yao2005,Hall2007,James2009,Yuan2010,Zhou2013,Lin2017, Shen2004,Zhang2011,Zhu2012, cao2020estimation,wang2020functional, shin2009partial,kong2016partially}, with a focus on estimation of the slope operator in one of these models.

It is also of practical importance to check whether the predictor $X$ has influence on the response $Y$ in the postulated model \eqref{FLM}, which corresponds to whether the slope operator is null and can be cast into the following hypothesis testing problem
\begin{equation}
	\mathrm{H}_{0}:\beta=0\quad\text{v.s.}\quad\mathrm{H}_{a}:\beta\neq0.	
	\label{HT-origin}
\end{equation}
This problem has been investigated in the literature, with more attention given to the scalar-on-function model. For example, among many others, \cite{Hilgert2013} proposed Fisher-type parametric tests with random projection to empirical functional principal components by using multiple testing techniques, \cite{Lei2014} introduced an exponential scan test by utilizing the estimator for $\beta$ proposed in \cite{Hall2007} that is based on functional principal component analysis, \cite{Qu2017} developed generalized likelihood ratio test using smoothing splines, and \cite{xue2021hypothesis},  exploiting the techniques developed for post-regularization inferences, constructed a test  for the case that there are an ultrahigh number of functional predictors.  For the function-on-function model, \cite{kokoszka2008testing} proposed a weighted $L_2$ test statistic based on functional principal component analysis{, and \cite{lai2021testing} developed a goodness-of-fit test  based on generalized distance covariance}. 
For the function-on-vector model, \cite{Shen2004, Zhang2011, smaga2019general} proposed functional F-tests while \cite{Zhu2012} considered a wild bootstrap method.


In this paper, we develop a novel approach to testing \eqref{HT-origin} under the model \eqref{FLM}, with 
the following distinct features. 
First, by exploiting principal component analysis of $X$ and $Y$, we propose a suitable transformation that transfers the test on the slope operator to a test on a high-dimensional vector of entries $\nu_{jk}:=\expect(\langle X-\expect X,\phi_j\rangle_1\langle Y-\expect Y,\psi_k\rangle_2)$, where $\phi_j$ and $\psi_k$ are population principal components of $X$ and $Y$, respectively; see Section \ref{sec:meth} for details. While there exist methods \citep[e.g.,][]{kokoszka2008testing,Hilgert2013,Lei2014,Su2017} that transfer testing \eqref{HT-origin} into testing  vectors, these vectors consist of the coefficients of $\beta$ with respect to $\phi_j$ and $\psi_k$. These coefficients however involve $\lambda_j^{-1}$ of the eigenvalues $\lambda_j$ that decay to zero in the setting of functional data, thereby facing the issue of ill-posedness. In contrast, the novelty of  our  transformation lies in eliminating $\lambda_{j}^{-1}$ and thus fully circumventing the ill-posed problem. In particular, it allows for incorporating all empirical principal components into our test statistic. Consequently, unlike existing counterparts, our test does not require an intricate choice of the number of empirical principal components, thereby enhancing  numeric stability and potentially increasing the test's power. 

Second, observing that the entries $\nu_{jk}$ are population means of the random variables $\langle X-\expect X,\phi_j\rangle_1\langle Y-\expect Y,\psi_k\rangle_2$, we propose to construct a max-type test statistic and bootstrap it via exploiting the inherent variance decay patterns of these random variables,  achieving higher power especially when the sample size is limited and/or the signal is weak. While the strategy of bootstrapping a max-type statistic  has been explored for  testing  the mean function \citep[e.g.,][]{Lopes2020,lin2020high}, it has not been studied in the substantially more challenging setting of functional linear regression. For example, in contrast to the works of \cite{Lopes2020,lin2020high}, which utilize a fixed and known projection basis, our test distinctively employs  \emph{empirical} principal components as a projection basis to construct the max-type test statistic. The integration of principal components and their empirical versions is crucial in our context, as the aforementioned transformation for fully circumventing ill-posedness relies explicitly on the principal components of $X$, while these principal components are unknown and practically estimated from data. Estimating these principal components  introduces nontrivial variability into the bootstrap procedure and substantially complicates the theoretical investigation; see Section \ref{sec:thm} for details.

In addition to the above methodological contributions, our theoretical studies not only  establish validity and consistency of the proposed test, but also uncover a practical and theoretical distinction between estimation and statistical inference about the slope operator $\beta$ when principal components are employed. For estimating the slope function, a delicate choice of the number $p$ of principal components is required to achieve consistent estimation \citep[e.g.,][]{Hall2007}; typically, $p\ll \sqrt n$ for data of $n$ observations. In sharp contrast, we show that, both numerically and theoretically, our test is valid and consistent even when all the $n$ principal components are included (i.e., $p=n$); note that at most $n$ principal components can be derived from $n$ observations. This not only eliminates the intricate tuning step for selecting the number of principal components, but also potentially improves numeric power of the test, particularly when the signal of the slope operator is tied to some high-order principal components. {This finding  highlights a critical distinction between estimation and inference about the slope operator/function in functional data analysis.} To the best of our knowledge, this is the first test utilizing all $p=n$ principal components. In contrast, at most $p\lesssim \sqrt{n}$ principal components are allowed by the previous studies in the general case \citep[e.g.,][]{cardot2003testing,muller2005generalized,Hilgert2013,Lei2014,Su2017,Choi2018,kokoszka2008testing,shin2009partial}.

In our theoretical investigation, to accommodate the situation that empirical principal components are adopted for conducting the aforementioned transformation, we establish validity and consistency of the proposed test uniformly for a family of test statistics induced by a class of orthonormal bases; we show that the empirical principal components fall into this class of bases with probability tending to one. We achieve  this by deriving \emph{uniform} Gaussian and bootstrap approximations of distributions of the corresponding family of max statistics. Consequently, our theoretical analyses are  materially different from, and encounter considerably more challenges than, the analyses in \cite{Lopes2020} and \cite{lin2020high}, which focus on only a single max statistic.
	For example, as discussed in Section \xref{sec:mean-model} of \cite{lin2024supphypothesis}, our analyses involve random elements in an infinite-dimensional Hilbert space, and thus the framework of  \cite{Lopes2020} for finite-dimensional Euclidean spaces does not directly apply. As another example, the orthonormal bases within the class, aside from the principal components $\phi_j$ and $\psi_k$, may destroy  the variance decay patterns induced by the principal components, which presents  a significant challenge to the techniques of \cite{Lopes2020,lin2020high}. Overcoming this challenge requires us to nontrivially and more effectively exploiting the basic property that $X$ and $Y$ have finite total variance, i.e.,  $\expect\langle X,X\rangle_1<\infty$ and $\expect\langle Y,Y\rangle_2<\infty$. These, along with other distinctions presented in our proofs, are materially different from the analysis in \cite{Lopes2020} and \cite{lin2020high}.

The rest of the paper is organized as follows. We describe the proposed test in Section \ref{sec:meth} and analyze its theoretical properties in Section \ref{sec:thm}. We then proceed to showcase its numerical performance via simulation studies in Section \ref{sec:simulation} and illustrate its applications in Section  \ref{sec:applications}. We conclude the article with a remark in Section \ref{sec:dis}. All proofs are provided in \cite{lin2024supphypothesis}.

\section{Methodology}\label{sec:meth}

Without loss of generality, we  assume $X$ and $Y$ in \eqref{FLM} are centered, i.e., $\expect X=0$ and $\expect Y=0$.  Such an assumption, adopted also in \cite{cai2006prediction}, is practically satisfied by replacing $X_i$ with $X_i - \bar{X}$ and replacing $Y_i$ with $Y_i - \bar{Y}$, where  $\bar{X}=n^{-1}\sum_{i=1}^n X_i$ and $\bar{Y}=n^{-1}\sum_{i=1}^n Y_i$. This simplifies the model \eqref{FLM} to 
\begin{equation}
	Y = \beta (X) + Z.
	\label{unified_equation}
\end{equation}
We assume $\expect \xnorm{X}^2<\infty$ and $\expect \ynorm{Y}^2<\infty$ {where $\xnorm\cdot$ and $\ynorm\cdot$ are norms induced respectively by $\xinnerprod{\cdot}{\cdot}$ and $\yinnerprod{\cdot}{\cdot}$}, so that the covariances of $X$ and $Y$ exist. 
Our goal is to test \eqref{HT-origin} based on the independently and identically distributed (i.i.d.) realizations $(X_1,Y_1),\ldots,(X_n,Y_n)$. 
In addition, we assume that $X_i$ and $Y_i$ are fully observed when they are functions. This assumption is pragmatically satisfied when $X_i$ and $Y_i$ are observed in a dense grid of their defining domains, as the observations in the grid can be interpolated to form an accurate approximation to $X_i$ and $Y_i$. 
{Thanks to modern technologies, such densely observed functional data are nowadays common in many fields, such as  medicine and healthcare \citep{Zhu2012,Chang2020+}, meteorology \citep{burdejova2017change, shang2017functional} and finance \citep{muller2011functional, tang2021forecasting}.}
The case that $X_i$ and $Y_i$ are only observed in a sparse grid is much more challenging and is left for future research.

For $x\in\mathcal X$ and $y\in\mathcal Y$, the tensor product operator $(x \otimes y): \mathcal X \rightarrow \mathcal Y$ is defined by 
\[
(x \otimes y)z = \langle x, z \rangle_1 y
\]
for all $z \in \mathcal{X}$. The tensor product $x\otimes z$ for $x,z\in\mathcal X$ is defined analogously. 
For example, if $\mathcal X=\real^q$, then $x \otimes z = z x^{\transpose}$ for $x,z\in\mathcal X$, and if $\mathcal X=L^2(\tdomain)$,  $f\otimes g$ is represented by the function $(f \otimes g)(s, t) = f(s)g(t)$, for $f,g \in L^{2}(\tdomain)$.
With the above notation, the covariance operator of a random element $X$ in the Hilbert space $\mathcal X$ is given by $C_{X} =  \expect (X\otimes X)$. 
For example, if $\mathcal{X}=\real^{q}$ then $C_X=\expect(XX^\transpose)$ and if $\mathcal{X}=L^{2}(\tdomain)$ then $(C_X f)(t)=\int_{\tdomain}\expect\{X(s)X(t)\}f(s)\diffop s$ for $f\in L^2(\tdomain)$ and all $t\in\tdomain$.

By Mercer's theorem, the operator $C_X$ admits the  decomposition
\begin{equation}\label{eq:CX}
	C_{X} = \sum_{j_1=1}^{d_{\mathcal X}} \lambda_{j_1} \phi_{j_1} \otimes \phi_{j_1},
\end{equation}
where $\lambda_1 > \lambda_2 > \cdots>0$ are eigenvalues, $\phi_1,\phi_2,\ldots$ are the corresponding eigenelements that are orthonormal, and $d_{\mathcal X}$ is the dimension of $\mathcal X$; for example, $d_{\mathcal X}=q$ if $\mathcal{X}=\real^q$ and $d_{\mathcal X}=\infty$ if $\mathcal X=L^2(\tdomain)$. 
Similarly, the operator $C_{Y}$ is decomposed by 
\begin{equation}\label{eq:CY}
	C_{Y} = \sum_{j_2=1}^{d_{\mathcal Y}} \rho_{j_2} \psi_{j_2} \otimes \psi_{j_2},
\end{equation}
with eigenvalues
$\rho_{1} > \rho_{2} > \cdots > 0$ and the corresponding eigenelements $\psi_1,\psi_2,\ldots$. Without loss of generality, we assume $\phi_1,\phi_2,\ldots$ form a complete orthonormal system (CONS) of $\mathcal X$ and $\psi_1,\psi_2,\ldots$ form a CONS of $\mathcal Y$; otherwise, we can simply redefine $\mathcal{X}$ and $\mathcal{Y}$ to be the closed subspaces spanned by the respective eigenelements corresponding to the nonzero eigenvalues, since $\beta$ in \eqref{unified_equation} can only be identified within the space of Hilbert--Schmidt operators between these subspaces.

Let $\mathfrak{B}_{HS}(\mathcal X, \mathcal Y)$ be the set of Hilbert--Schmidt operators from $\mathcal X$ to $\mathcal Y$ \citep[see Definition 4.4.2 in][]{Hsing2015}, 
and note that $\beta \in \mathfrak{B}_{HS}(\mathcal{X}, \mathcal{Y})$. Since $\phi_1,\phi_2,\ldots$ and $\psi_1,\psi_2,\ldots$ are CONS,  $\beta$ can be represented as 
\begin{equation}
	\beta = \sum_{j_1=1}^{d_{\mathcal X}} \sum_{j_2=1}^{d_{\mathcal Y}} b_{j_1j_2} \phi_{j_1} \otimes \psi_{j_2},
	\label{beta-expansion}
\end{equation}
where each $b_{j_1j_2} \in \mathbb{R}$ is the generalized Fourier coefficient. Consequently, the null hypothesis in \eqref{HT-origin} is equivalent to $b_{j_1j_2}=0$ for all $j_1$ and $j_2$.  
It turns out that the coefficients are linked to the cross-covariance operator $\expect (X \otimes Y)$. Specifically, with $\nu_{j_1j_2} = \langle \expect (X \otimes Y), \phi_{j_1} \otimes \psi_{j_2} \rangle$, we have the following proposition that connects $b_{j_1j_2}$ and $\nu_{j_1,j_2}$; special cases of this connection have been exploited for example by \cite{cai2006prediction,Hall2007,kokoszka2008testing}. 

\begin{proposition}\label{prop:coef-relation}
	$\nu_{j_1 j_2} = \expect ( \langle X, \phi_{j_1} \rangle_1 \langle Y, \psi_{j_2} \rangle_2 )$ and $b_{j_1j_2} = \lambda_{j_1}^{-1} \nu_{j_1j_2}$.
\end{proposition}

{Because $\lambda_{j_1}\rightarrow 0$ as $j_1\rightarrow\infty$, estimating the coefficients $b_{j_1j_2}$ becomes an ill-posed problem \citep{Hall2007} and hence a direct test on the coefficients is difficult.}
To overcome the challenge of ill-posedness, we go one step further to observe that, $b_{j_1j_2}=0$ is  equivalent to $\nu_{j_1j_2}=0$ for all $j_1$ and $j_2$, as the eigenvalues $\lambda_{j_1}$ are  nonzero according to remark right after \eqref{eq:CY}. Therefore, a test on $b_{j_1j_2}$ can be further transformed into a test on $\nu_{j_1j_2}$, and this consequently eliminates the difficulty of estimating the reciprocals of the eigenvalues and the associated complications in deriving the asymptotic distribution of the test statistic. This further transformation, previously not exploited in the literature, is elegantly simple and effective for fully circumventing ill-posedness of functional linear regression in the context of testing nullity of the slope operator. 
Moreover, testing $\nu_{j_1j_2}=0$  is much more manageable as each $\nu_{j_1j_2}$ is the mean of some random variable according Proposition \ref{prop:coef-relation}.

Following from the above discussions, we consider testing $\nu_{j_1j_2}=0$ for $j_1=1,\ldots,p_1$  when $d_{\mathcal X}=\infty$, and analogously, for $j_2=1,\ldots,p_2$ when $d_{\mathcal Y}=\infty$. Here, $p_1$ and $p_2$ are integers that may grow with the sample size; when $d_{\mathcal X}<\infty$ or $d_{\mathcal Y}<\infty$, one may opt for $p_1=d_{\mathcal X}$ or $p_2=d_{\mathcal Y}$, respectively. 
Given $p_1$ and $p_2$, we focus on the truncated vector $\nu=(\nu_{j_1j_2}: (j_1,j_2)\in \P)$ comprising $p=p_1p_2$ elements, with $\P=\{(j_1,j_2): j_1=1,\ldots, p_1, j_2=1, \ldots, p_2\}$. 
Formally, with this configuration of $\nu$, we pragmatically consider the following surrogate hypothesis testing problem
\begin{equation}\label{eq:trunc-ht}
	\mathrm{H}_0: \nu=0
	\qquad\text{versus}\qquad 
	\mathrm{H}_a: \nu\ne 0,
\end{equation}

To test the above hypothesis, 
we observe that the random variables $\xinnerprod{X}{\phi_{j_1}}\yinnerprod{Y}{\psi_{j_2}}$ have the expected values $\nu_j$ and their variances exhibit a decay pattern under some regularity conditions; {see Section \ref{sec:thm} for details}. This motivates us to adapt the idea of partial standardization developed in \cite{Lopes2020,lin2020high}. Specifically, we consider the following max and min statistics and their  asymptotic distributions,
\begin{equation}\label{eq:M}
	M:=M(\phi,\psi)=\underset{1\leq j\leq p}{\max}\frac{S_{n,j}}{\sigma_j^{\tau}}\qquad\qquad L:=L(\phi,\psi)=\underset{1\leq j\leq p}{\min}\frac{S_{n,j}}{\sigma_j^{\tau}},
\end{equation} 
where $\tau\in[0,1)$ is a tuning parameter, $S_{n,j}$ is the $j$th coordinate of $S_n:=n^{-1/2}\sum_{i=1}^n \{V_i-\nu\}$ with $V_i$ being the vector formed by $\xinnerprod{X_i}{\phi_{j_1}}\yinnerprod{Y_i}{\psi_{j_2}}$ for $(j_1,j_2)\in \P$, and $\sigma_j^2=\mathrm{var}(V_{ij})$ with $V_{ij}$ being the $j$th coordinate of $V_i$. Intuitively, $\max_{1\leq j\leq p} n^{-1/2}\sum_{i=1}^nV_{ij}/\sigma_j^\tau$ has the same distribution with $M$ under $\mathrm{H}_0$ and may be much larger than $M$ under $\mathrm{H}_a$; similar intuition applies to the random quantity $\min_{1\leq j\leq p}n^{-1/2}\sum_{i=1}^nV_{ij}/\sigma_j^\tau$. This leads us to the following test statistics
$$ T_U = \underset{1\leq j\leq p}{\max}\frac{\sqrt n\bar V_j}{\hat\sigma_j^\tau}\quad\text{ and }\quad T_L=\underset{1\leq j\leq p}{\min}\frac{\sqrt n\bar V_j}{\hat\sigma_j^\tau},$$
where $\bar V_j$ represents the $j$th coordinate of  $\bar V=n^{-1}\sum_{i=1}^n V_i$, and $\hat \sigma^2_j$, which is an estimate of $\sigma_j^2$, is the $j$th diagonal element of $\hat\Sigma=n^{-1}\sum_{i=1}^n(V_i-\bar V)(V_i-\bar V)^{\transpose}$. For a significance level $\varrho$, we may reject the null hypothesis if $T_U$ exceeds its $1-\varrho/2$ quantile or $T_L$ is below its $\varrho/2$ quantile.

It remains to estimate the quantiles of $T_U$ and $T_L$ for any given $\varrho\in(0,1)$ under the null hypothesis, for which we adopt a bootstrap strategy, as follows. Let $S^\star_n$ be drawn from the distribution $N_p(0,\hat\Sigma)$ conditional on the data, where $N_p(0,\hat\Sigma)$ denotes the $p$-dimensional centered Gaussian distribution with the covariance matrix $\hat\Sigma$. Then, the bootstrap counterparts of $M$ and $L$ are given by
$$M^\star = \underset{1\leq j\leq p}{\max}\frac{S_n^\star}{\hat\sigma_j^\tau}\quad\text{ and }\quad L^\star = \underset{1\leq j\leq p}{\min}\frac{S_n^\star}{\hat\sigma_j^\tau},$$
respectively. Intuitively, the distribution of $M^\star$ provides an approximation to the distribution of $M$ when the sample size is sufficiently large, while the distribution of $M$ acts as a surrogate for the distribution of $T_U$ under $H_0$; we justify this intuition in Theorems \ref{thm:gaussian-app} and \ref{thm:bootstrap-app}.
{Therefore, we 
	$$\text{
		reject the null hypothesis if  }T_{U} > q_{M^{\star}}(1-\varrho/2)  \text{ or }T_{L} < q_{L^{\star}}(\varrho/2), $$
	where $q_{M^{\star}}(\cdot)$ and $q_{L^{\star}}(\cdot)$ are quantile functions of $M^{\star}$ and $L^{\star}$ respectively. 
	In particular, both quantile functions can be practically approximated via resampling from the distribution $N_p(0,\hat\Sigma)$.} Specifically, for a sufficiently large integer $B$, e.g., $B=1000$, for each $b=1,\ldots,B$, we independently draw $S_{n}^{\star,b}\sim N_p(0,\hat\Sigma)$ and compute $M^{\star,b}$ and $L^{\star,b}$. The quantiles {$q_{M^{\star}}(1-\varrho/2)$ and $q_{L^{\star}}(\varrho/2)$ are then} respectively estimated by the empirical $1-\varrho/2$ quantile $\hat q_{M}(1-\varrho/2)$ of $M^{\star,1},\ldots,M^{\star,B}$ and the  $\varrho/2$ quantile $\hat q_L(\varrho/2)$ of $L^{\star,1},\ldots,L^{\star,B}$.

In practice, the eigenelements $\phi_{j_1}$ and $\psi_{j_2}$ are unknown. To test \eqref{eq:trunc-ht}, we need to estimate $\phi_{j_1}$ and $\psi_{j_2}$ from data. Specifically, $\phi_{j_1}$ is estimated by the eigenelement corresponding to the $j_1$th eigenvalue of the sample covariance operator $\hat C_X=n^{-1}\sum_{i=1}^n X_i\otimes X_i$, and similarly, $\psi_{j_2}$ is estimated by the eigenelement corresponding to the $j_2$th eigenvalue of $\hat C_Y=n^{-1}\sum_{i=1}^n Y_i\otimes Y_i$. 
The tuning parameter $\tau\in[0,1)$ controls the degree of partial standardization in \eqref{eq:M} and is the key to exploiting the decay variance. The work of \cite{lin2020high} provides a strategy to select a value of $\tau$ that  maximizes the empirical power of the test, where  the projection basis is fixed and known. In our numeric studies presented in Section \ref{sec:simulation}, we found that the same selection strategy empirically works well even when the projection bases are estimated from data in our case.

As aforementioned in the introduction, in the previous studies of functional linear regression, it is crucial to determine a delicate value for  $p_1$ and $p_2$ (typically  $\lesssim\sqrt{n}$)  to ensure estimation consistency or test validity. In contrast,  our theoretical results in the next section show that our test remains asymptotically valid and consistent even with the choice of $p_1=p_2=n$; note that at most $n$ empirical principal components can be derived from $n$ observations. This choice of $p_1$ and $p_2$ is also validated by our numeric studies in Section \ref{sec:simulation}. Consequently, unlike the existing counterparts, our test eliminates the nontrivial requirement of selecting the number of principal components, leading to enhanced numeric simplicity and stability.

\section{Theory}\label{sec:thm}

We begin with introducing some notations. The symbol $\ell^2$ denotes the set of sequences that are square summable.
For a matrix $A$, we write $\|A\|_{F}=\left(\sum_{i,j} A_{ij}^{2}\right)^{1/2}$ for its Frobenius norm and $\|A\|_{\infty} = \max_{i,j} |A_{ij}|$ for its max norm, where $A_{ij}$ is the element of $A$ at position $(i,j)$. 
{For a random variable $\xi$ and a real number $\theta\in(0,2]$, the $\psi_{\theta}$-Orlicz (quasi-)norm is defined as  $\|\xi\|_{\psi_{\theta}}=\inf\{t>0:\expect[\exp(|\xi/t|^{\theta})]\le 2\}$, where the  cases of $\theta=1$ and $\theta=2$ correspond to the sub-exponential and sub-Gaussian random variables, respectively.} 
We also use $\mathcal{L}(\xi)$ to denote the distribution of $\xi$ and define the Kolmogorov distance between random variables $\xi$ and $\eta$ by  $d_{\text{K}}\left(\mathcal{L}(\xi), \mathcal{L}(\eta)\right)=\sup_{t\in \real}\left| \mathbb{P}(\xi\le t) - \mathbb{P}(\eta \le t) \right|$. 
For two sequences $\{a_n\}$ and $\{b_n\}$ with non-negative elements, $a_n = o(b_n)$ or $a_n \ll b_n$ means $a_n / b_n \to 0$ as $n \to \infty$, and $a_n = O(b_n)$ means $a_n \le c b_n$ for some constant $c>0$ and all sufficiently large  $n$. Moreover, we write $a_n \lesssim b_n$ if $a_n=O(b_n)$, write $a_n \gtrsim b_n$ if $b_n=O(a_n)$, and write $a_n \asymp b_n$ if $a_n \lesssim b_n$ and $a_n \gtrsim b_n$. Also, define $a_n \vee b_n = \max\{a_n, b_n\}$ and $a_n \wedge b_n = \min\{a_n, b_n\}$.

%
Consider the untruncated vector $V_{i}^{\infty}=(\xi_{ij_1} \zeta_{ij_2}, j_1,j_2\ge 1)$ with $\xi_{ij_1}=\langle X_i, \phi_{j_1} \rangle_1$ and $\zeta_{ij_2}=\langle Y_i, \psi_{j_2} \rangle_2$. Our first assumption concerns the tail behavior of $V_i^\infty$.

\begin{assumption}[Tail behavior]\label{as:tail}
	The random vector $V_{1}^{\infty}$ satisfies the following two tail conditions: 
	\begin{align}
		&\prob\left(\|V_{1}^{\infty}-\expect V_{1}^{\infty}\|_2\ge t\right) \le 2 e^{-\frac{t}{K}} \label{cond-nSE} \\
		&\| \langle V_{1}^{\infty}-\expect V_{1}^{\infty}, x \rangle \|_{\psi_1} \le K \Big(\expect\big[\langle V_{1}^{\infty}-\expect V_{1}^{\infty}, x \rangle^2\big]\Big)^{1/2} \label{cond-SE-conc}
	\end{align}
	for some $K>0$, all $t\ge 0$, and all $x\in \ell^2$ with $\|x\|_2=1$.
\end{assumption}


Condition \eqref{cond-nSE}, extending \cite{jin2019short} to sub-exponentiality and potentially infinite-dimensional vectors, ensures that the $\ell_2$-norm of the centered $V_1^\infty$ has a sub-exponential tail. {For example, it holds when both $X$ and $Y$ have sub-Gaussian norms, i.e., when $\|X\|_1$ and $\|Y\|_2$ are sub-Gaussian.}  
Condition \eqref{cond-SE-conc}, e.g., satisfied by normal random elements and vectors, extends its benchmark counterparts in  \cite{Lopes2020,antonini1997subgaussian, vershynin2018high, giessing2020bootstrapping,cai2022sparse} to the infinite-dimensional setting. 

{To state the next assumption, let $R(d_1, d_2)\in \real^{d_1d_2\times d_1d_2}$ denote the correlation matrix of random variables $\{\langle X,\phi_{j_1}\rangle_1\langle Y,\psi_{j_2}\rangle_2: 1\leq j_1\leq d_1,~1\leq j_2\leq d_2\}$. That is, $R(d_1, d_2)$ corresponds to the cross-products of the leading $d_1$ principal component scores of $X_1$ and the leading $d_2$ principal component scores of $Y_1$.}

\begin{assumption}[Structural assumptions]\label{as:structure}
	~
	\begin{enumerate}[label=(\roman*)]
		\item 
		{$\expect[\langle X_1, \phi_{j_1} \rangle_1^4]\le C \lambda_{j_1}^2$ for some $C>1$ and all $j_1\ge 1$}. 
		The eigenvalues $\lambda_{j_1}$ for $j_1\ge 1$ and $\rho_{j_2}$ for $j_2\ge 1$ are positive, and there are constants {$\alpha_1>2$} and  $\alpha_2>1$, not depending on $n$, such that
		\begin{equation}
			\lambda_{j_1} \asymp {j_1}^{-\alpha_1} \quad \text{and} \quad \rho_{j_2} \asymp {j_2}^{-\alpha_2}.
			\label{eigen-rate}
		\end{equation}
		Moreover,
		\begin{equation}
			\max_{j_1 \ge 1} |b_{j_1 j_2}| \lesssim \rho_{j_2}, \quad \text{for all}~ j_2\ge 1.
			\label{beta-alignment}
		\end{equation}
		\item Let $\bar{\alpha}=\max\{{\alpha_1/2}, \alpha_2\}$ and $\underline{\alpha}=\min\{{\alpha_1/2}, \alpha_2\}$.	For  any constant $\delta \in (0, 1/2)$ and an arbitrarily small number $\delta_{0}>0$, define $\ell_{n} = \lceil n^{\frac{\delta}{4 \vee (3 \underline{\alpha} (1-\tau))}} \wedge p \rceil$, $m_n=\ell_n^{\frac{2\bar{\alpha}}{\underline{\alpha}-1}+\delta_0}$ and the class
		\[
		\mathcal{R}(\ell_{n}, m_{n}) = \left\{ R^{\circ} \in \real^{\ell_{n} \times \ell_{n}}: R^{\circ} \text{ is a sub-matrix of } R(\lfloor m_{n}^{1/2}\rfloor,\lfloor m_{n}^{1/2}\rfloor) \right\}.
		\]
		Assume
		\[
		\sup_{R^{\circ} \in \mathcal{R}(\ell_{n}, m_{n})}\|R^{\circ}\|_{F}^{2} \lesssim \ell_{n}^{2-\delta}.
		\]
	\end{enumerate}
\end{assumption}

The condition on $\expect[\langle X_1, \phi_{j_1} \rangle_1^4]$  was previously adopted in \cite{Cai2006, Hall2007, Lei2014}. 
The condition \eqref{eigen-rate} imposes a smoothness requirement on the covariance operators of $X$ and $Y$, which is often needed in analyzing convergence rates involving functional data; for example, similar requirements are adopted in \cite{meister2011asymptotic, cai2018adaptive}. Such condition is connected to the so-called Sacks-Ylvisaker condition \citep{yuan2010reproducing}. For example, {
	when the covariance $C_X$ satisfies the Sacks-Ylvisaker condition of order $s > 0$, the corresponding $j$th eigenvalue is of the order $j^{-2(s+1)}\ll j^{-2}$.}
For a scalar-on-function model, $p_2=d_{\mathcal Y}=1$, and thus the requirement for $\rho_{j_2}$ in \eqref{eigen-rate} and the condition on the generalized Fourier coefficients $b_{j_1j_2}$ of $\beta$ in \eqref{beta-alignment} are automatically satisfied. In addition, \eqref{beta-alignment} is considerably weaker than the requirement in \cite{cai2006prediction, Hall2007} for the  scalar-on-function regression model.
Condition (ii) in the above assumption is in analogy to Condition (3.2) of \cite{lopes2021sharp}. As mentioned in \cite{lopes2021sharp}, since $\sup_{R^{\circ} \in \mathcal{R}(\ell_{n}, m_{n})}\|R^{\circ}\|_{F}^{2}\le \ell_n^2$ and   $\delta$ could be taken arbitrarily small, the condition (ii), which cannot be substantially weakened in general, appears not restrictive. {In addition, this condition applies only to the small set of the variables that involve the $m_n$ leading eigen-bases, while the other variables can  have an unrestricted correlation structure.} 

Our last assumption imposes some conditions on the growth rate of $p$ relative to $n$, where we recall that $p=p_1p_2$.

\begin{assumption}\label{as:tech}  
	We require $p \lesssim n^{\alpha_0}$ for a (arbitrarily large) fixed number $\alpha_0>0$. 
\end{assumption}

Under this assumption, $p_1$ and $p_2$, the numbers of potential scalar predictors or principal component scores, are allowed to grow with the sample size at a polynomial  rate of $n$. Hence, in the scenario where $d_{\mathcal X}=\infty$ and/or $d_{\mathcal Y}=\infty$, the assumption accommodates the situation where $p_1=n$ and/or $p_2=n$, which corresponds to the maximal number of  empirical eigenfunctions that can be derived from $n$ observations.
This is markedly distinct from estimation problems that involve empirical eigenfunctions \citep[e.g.,][]{Hall2007}, in which at most $O(\sqrt n)$ leading eigenfunctions can be utilized in order to guarantee consistent estimation of $\beta$ and the eigenfunctions \citep[e.g.,][]{wahl2022lower}. In contrast, for testing  the slope function, $p_1=n$ and/or $p_2=n$ are allowed,  as  demonstrated by the proposed test procedure. This  suggests that, for statistical inference about the slope function in functional data analysis, we may include all $n$ empirical eigenfunctions, without requiring consistent estimation of {all} eigenfunctions being involved. 


As mentioned previously, the eigenelements $\phi=\{\phi_{j_1}\}_{j_1=1}^{p_1}$ and $\psi=\{\psi_{j_2}\}_{j_2=1}^{p_2}$ are often unknown, and practitioners may use alternative orthonormal elements $\tilde\phi=\{\tilde\phi_{j_1}\}_{j_1=1}^{p_1}$ and $\tilde\psi=\{\tilde\psi_{j_2}\}_{j_2=1}^{p_2}$, such as the empirical eigenelements $\hat\phi=\{\hat\phi_{j_1}\}_{j_1}^{p_1}$ and $\hat\psi=\{\hat\psi_{j_2}\}_{j_2}^{p_2}$,  which may differ from $\phi$ and $\psi$. {In this case, all quantities depending on $\phi$ and $\psi$, such as $M$ and $S_n$, will be computed by using $\tilde\phi$ and $\tilde\psi$. We write, for example, $M(\tilde\phi,\tilde\psi)$ and $S_n(\tilde\phi,\tilde\psi)$, to indicate the dependence on $\tilde\phi$ and $\tilde\psi$, and note that $M=M(\phi,\psi)$ and $S_n=S_n(\phi,\psi)$.}

For two integers $d_1, d_2\ge 1$, for two orthonormal sequences $\{\tilde{\phi}_{j_1}\}_{j_1=1}^{d_1}$ and $\{\tilde{\psi}_{j_2}\}_{j_2=1}^{d_2}$, define 
\begin{equation*}
	U_{\mathcal{X}}^{d_1}(\tilde{\phi}) = 
	\begin{pmatrix}
		\langle \phi_1, \tilde{\phi}_1 \rangle_1 & \cdots &  \langle \phi_1, \tilde{\phi}_{d_1} \rangle_1 \\
		\vdots  & \ddots & \vdots  \\
		\langle \phi_{d_1}, \tilde{\phi}_1 \rangle_1 & \cdots &  \langle \phi_{d_1}, \tilde{\phi}_{d_1} \rangle_1
	\end{pmatrix}
	\quad \text{and} \quad
	U_{\mathcal{Y}}^{d_2}(\tilde{\psi}) =  
	\begin{pmatrix}
		\langle \psi_1, \tilde{\psi}_1 \rangle_2 & \cdots &  \langle \psi_1, \tilde{\psi}_{d_2} \rangle_2 \\
		\vdots  & \ddots & \vdots  \\
		\langle \psi_{d_2}, \tilde{\psi}_1 \rangle_2 & \cdots &  \langle \psi_{d_2}, \tilde{\psi}_{d_2} \rangle_2
	\end{pmatrix}.
\end{equation*}
Let $W_d(\tilde{\phi}, \tilde{\psi}) = U_{\mathcal{X}}^{d_1}(\tilde{\phi}) \kronprod U_{\mathcal{Y}}^{d_2}(\tilde{\psi})$, where $\kronprod$ denotes the Kronecker product of two matrices, be the transformation matrix consisting of the leading eigenelements.
With the truncation numbers $p_1$ and $p_2$ introduced in Section \ref{sec:meth}, consider a class $\mathcal F_p$ of $(\tilde\phi,\tilde\psi)$ with $\tilde\phi=\{\tilde\phi_{j_1}\}_{j_1=1}^{p_1}$ and  $\tilde\psi=\{\tilde\psi_{j_2}\}_{j_2=1}^{p_2}$, such that 
1) $\tilde\phi$ and $\tilde\psi$ are respectively orthonormal sequences, 
and {2) 
	\begin{equation}
		\max\left\{\|U_{\mathcal{X}}^{d_1}(\tilde{\phi})-I_{d_1}\|_{\infty}, \|U_{\mathcal{Y}}^{d_2}(\tilde{\psi})-I_{d_2}\|_{\infty}, \| W_{d}(\tilde{\phi}, \tilde{\psi}) - {I}_{d} \|_{\infty}\right\} \leq c a_{n}
		\label{eq:Fp-transmat-require}
	\end{equation}
	for a sufficiently large constant $c>0$, 
	{where ${I}_{r}$ represents the $r\times r$ identity matrix for any integer $r>0$, $d=d_1d_2$, and we set
		\begin{align*}
			a_n & =k_n^{-2\bar{\alpha}}\quad\quad\text{with}\quad k_n=\ell_n^{\frac{(4\alpha_0+2)\bar\alpha}{\min\{1, \underline{\alpha}-1\}}}, \\
			d_1 & =  \min\{\lfloor \Tn^{1/2}\rfloor, p_1\} \quad\text{and}\quad d_2=\min\{\lfloor \Tn^{1/2}\rfloor, p_2\} \qquad\text{with}\quad \Tn=n^{\frac{1}{2(\bar{\alpha}+1)}}
		\end{align*}
		throughout this paper. Note that $d\lesssim \Tn$, with the above choices. 

Th condition $\eqref{eq:Fp-transmat-require}$, for example, is satisfied by the leading empirical eigenbases with probability tending to one, according to the following proposition. 

\begin{proposition}\label{prop:empirical-eigenfunctions}
	Let $\{\hat{\phi}_{j_1}\}_{j_1=1}^{d_1}$ and $\{\hat{\psi}_{j_2}\}_{j_2=1}^{d_2}$ be the leading empirical eigenelements of $\hat{C}_{X}$ and $\hat{C}_{Y}$ defined in Section \ref{sec:meth}, respectively.  If $X$ and $Y$  are sub-Gaussian random elements in $\mathcal{X}$ and $\mathcal{Y}$ satisfying Assumption \ref{as:structure}, then for $1\ll t \ll \max\{d_1^{-2(\alpha_1+1)}n,d_2^{-2(\alpha_2+1)}n\}$,  with probability at least $1-e^{-t + 2\log(2d)}$, we have
	\[
	\max\left\{\|U_{\mathcal{X}}^{d_1}(\hat{\phi})-I_{d_1}\|_{\infty}, \|U_{\mathcal{Y}}^{d_2}(\hat{\psi})-I_{d_2}\|_{\infty}, \| W_d(\hat{\phi}, \hat{\psi}) - {I}_d \|_{\infty}\right\}
	\lesssim (d_1^{\alpha_1+1}\vee d_2^{\alpha_2+1}) \sqrt{t/n}.
	\]
	Consequently, {for any $q> 0$ such that 
		$t \asymp n/(k_n^{2q} \bar{d}^{2(\bar{\alpha}+1)}) \gg \log d$ with $\bar{d}=\max\{d_1, d_2\}$, we have
		\[
		\max\left\{\|U_{\mathcal{X}}^{d_1}(\hat{\phi})-I_{d_1}\|_{\infty}, \|U_{\mathcal{Y}}^{d_2}(\hat{\psi})-I_{d_2}\|_{\infty}, \| W_d(\hat{\phi}, \hat{\psi}) - {I}_d \|_{\infty}\right\} \lesssim k_n^{-q}
		\]
		with probability tending to one. In particular, it holds for $q=2\bar\alpha$.}
\end{proposition}

\begin{rem}\label{rem:p=n}
	In the case $d_{\mathcal X}=d_{\mathcal Y}=\infty$ so that in practice we set $p_1=p_2=n$, 
	{the requirement \eqref{eq:Fp-transmat-require} in $\mathcal{F}_p$, with the choice of $d_1$ and $d_2$, is imposed only on the leading $\lfloor \Tn^{1/2}\rfloor\ll p_1$ eigen-bases of $C_X$ and the leading $\lfloor \Tn^{1/2}\rfloor\ll p_2$ eigen-bases of $C_Y$.} In particular, there are no conditions on the remaining eigen-bases. 
	The condition \eqref{eq:Fp-transmat-require} would enforce that the variances of the coordinates of $V_i(\tilde\phi,\tilde\psi)$ partially exhibit a decay pattern similar to that of the variances of $V_i$; {see Proposition \xref{prop-struct-tilde-V} in \cite{lin2024supphypothesis} for details.}  This is one of the key properties we exploit for establishing the validity and consistency of the proposed test even when we take $p_1=p_2=n$ that is set in our numeric implementation. Another key property we heavily exploit is that $X$ and $Y$ have finite total variances, i.e., $\expect \|X-\expect X\|_1^2<\infty$ and $\expect \|Y-\expect Y\|_2^2<\infty$ or equivalently $\sum_{j_1=1}^\infty \lambda_{j_1}<\infty$ and $\sum_{j_2=1}^\infty \rho_{j_2}<\infty$.
\end{rem}

The class $\mathcal F_p$  gives rise to a class of test statistics $T_U(\tilde\phi,\tilde\psi)$ and $T_L(\tilde\phi,\tilde\psi)$.
Below we analyze the uniform asymptotic power and size over this class of test statistics; the asymptotic properties of the proposed test by using $T_U=T_U(\phi,\psi)$ and $T_L=T_L(\phi,\psi)$, as well as their empirical versions $T_U(\hat\phi,\hat\psi)$ and $T_L(\hat\phi,\hat\psi)$, then follow as direct consequences, since the class $\mathcal F_p$ contains $(\phi,\psi)$ and further $(\hat\phi,\hat\psi)$ with probability tending to one according to Proposition \ref{prop:empirical-eigenfunctions}. To this end, we  first establish three approximation results related to the test statistics, namely, the Gaussian approximation, the bootstrap approximation and the approximation with empirical variances, \emph{uniformly} over the class $\mathcal F_p$. Compared with their non-uniform counterparts in \cite{Lopes2020, lin2020high, lopes2021sharp}, these general uniform approximations, maybe of independent interest, require considerably different and more challenging proofs. For example, as mentioned in Remark~\ref{rem:p=n}, we heavily exploit the property that $X$ and $Y$ have finite total variances in our proofs, which is materially different from the aforementioned previous works. 
Below we consider only the max statistic while note that similar results hold for the min statistic. 

We start with defining the Gaussian counterpart of $M(\tilde\phi,\tilde\psi)$ by 
\[
\check{M}(\tilde{\phi},\tilde{\psi}) = \max_{1\leq j\leq p}\frac{\check{S}_{n,j}(\tilde{\phi},\tilde{\psi})}{\sigma_{j}^{\tau}(\tilde{\phi},\tilde{\psi})},
\]
where $\check{S}_n(\tilde{\phi},\tilde{\psi}) \sim N_p(0, \Sigma(\tilde{\phi},\tilde{\psi}))$.
The following result shows that the distribution of $M(\tilde{\phi},\tilde{\psi})$ converges to the distribution $\check M(\tilde{\phi},\tilde{\psi})$ at a near $1/\sqrt{n}$ rate uniformly over $\mathcal F_p$.

\begin{theorem}[Uniform Gaussian approximation]\label{thm:gaussian-app}
	For any sufficiently small number $\delta \in (0, 1/2)$, if Assumptions \ref{as:tail}--\ref{as:tech} hold, then
	\[
	\sup_{(\tilde{\phi}, \tilde{\psi}) \in \mathcal{F}_p}
	d_{\text{K}}\left(\mathcal{L}(M(\tilde{\phi},\tilde{\psi})), \mathcal{L}(\check{M}(\tilde{\phi},\tilde{\psi}))\right) \lesssim n^{-1/2 + \delta}.
	\]
\end{theorem}

In the proposed test, a bootstrap strategy is used to estimate the distribution of $\check M(\tilde{\phi},\tilde{\psi})$, which is justified by the following result.

\begin{theorem}[Uniform bootstrap approximation]
	For any sufficiently small number $\delta \in (0, 1/4)$, if Assumptions \ref{as:tail}--\ref{as:tech} hold, then there is a constant $c>0$, not depending on $n$, such that the event 
	\[
	\sup_{(\tilde{\phi}, \tilde{\psi}) \in \mathcal{F}_p}
	d_{\text{K}}\left(\mathcal{L}(\check{M}(\tilde{\phi},\tilde{\psi})), \mathcal{L}(M^{\star}(\tilde{\phi},\tilde{\psi})|\mathcal{D})\right) \leq cn^{-1/4 + \delta}
	\]
	occurs with probability at least $1-cn^{-1}$, where $\mathcal{L}(M^{\star}(\tilde{\phi},\tilde{\psi})|\mathcal{D})$ represents the distribution of $M^{\star}(\tilde{\phi},\tilde{\psi})$ conditional on the observed data $\mathcal{D} = (X_i, Y_i)_{i=1}^n$. 
	\label{thm:bootstrap-app}
\end{theorem}

\begin{rem}
	The rate of near $n^{-1/4}$ in the bootstrap approximation is primarily due to the requirement of \emph{uniform} convergence over $\mathcal F_p$ of growing complexity.
	Specifically, 
	for this uniform convergence, we need to establish a uniform Gaussian-to-Gaussian comparison result between the Gaussian and bootstrap counterparts of the max statistic over the $k_n$ leading components; details can be found in Lemma \xref{lemma:Wop} of \cite{lin2024supphypothesis}. 
	Our strategy is to transform these leading components induced by $(\tilde{\phi}, \tilde{\psi})$ into the components  induced by the population eigenelements $(\phi, \psi)$. 
	This transformation involves an infinite-dimensional operator, destroying some key finite-dimensional structures possessed  in the analysis of \cite{Lopes2020}, such as Proposition C.1 therein. Instead, we apply a Gaussian comparison result for general covariance structures, such as Lemma \xref{lemma:Gaussian-comparison-optimal} in \cite{lin2024supphypothesis}, leading to the rate $n^{-1/4+\delta}$.
\end{rem}

In reality, the variances $\sigma_j^2$ are estimated by $\hat\sigma_j^2$, and the max statistic is pragmatically computed by \[
\hat{M}(\tilde{\phi},\tilde{\psi}) = \max_{1\le j \le p} \frac{S_{n,j}(\tilde{\phi},\tilde{\psi})}{\hat{\sigma}_{j}^{\tau}(\tilde{\phi},\tilde{\psi})}.
\]
Below we show that the distribution of this practical max statistic converges to the distribution of the original max statistic uniformly over the class $\mathcal F_p$.

\begin{theorem}
	For any sufficiently small number $\delta \in (0, 1/2)$, if Assumptions \ref{as:tail}--\ref{as:tech} hold, then
	\[
	\sup_{(\tilde{\phi}, \tilde{\psi}) \in \mathcal{F}_p}
	d_{\text{K}}\left(\mathcal{L}(\hat{M}(\tilde{\phi},\tilde{\psi})), \mathcal{L}(M(\tilde{\phi},\tilde{\psi}))\right) \lesssim n^{-1/2 + \delta}.
	\]
	\label{thm:size-app}
\end{theorem}

With the triangle inequality, 
Theorems \ref{thm:gaussian-app}--\ref{thm:size-app} together imply that, with probability at least $1 - cn^{-1}$, 
\[
\sup_{(\tilde{\phi}, \tilde{\psi}) \in \mathcal{F}_p}
d_{\text{K}}\left(\mathcal{L}(\hat{M}(\tilde{\phi},\tilde{\psi})) , \mathcal{L}(M^{\star}(\tilde{\phi},\tilde{\psi})|\mathcal{D})\right) \le c n^{-1/4 + \delta}
\] 
for some constant $c>0$ not depending on $n$. 
This eventually leads to Theorem \ref{thm:size} and Theorem \ref{thm:power} for uniform validity and consistency of the proposed test respectively.

\begin{theorem}
	For any sufficiently small number $\delta \in (0, 1/4)$, if Assumptions \ref{as:tail}--\ref{as:tech} hold, then  for any $\varrho\in(0,1)$,  we have 
	\[
	\sup_{(\tilde{\phi},\tilde{\psi}) \in \mathcal{F}_p} \mathrm{SIZE}(\varrho, \tilde{\phi}, \tilde{\psi}) \leq \varrho  +  O(n^{-1/4+\delta}),
	\]
	{where $\mathrm{SIZE}(\varrho, \tilde{\phi}, \tilde{\psi})$ is the probability of rejecting the null hypothesis by using the bases $(\tilde\phi,\tilde\psi)$ at the significance level $\varrho$ when the null hypothesis in \eqref{HT-origin} is true.} 
	\label{thm:size}
\end{theorem}

\begin{theorem}Suppose Assumptions \ref{as:tail}--\ref{as:tech} hold. Then,
	\begin{enumerate}[label=\textup{(\arabic*)}]
		\item for any fixed $\varrho \in (0,1)$, one has
		\[
		\sup_{(\tilde{\phi},\tilde{\psi}) \in \mathcal{F}_p}
		|q_{M^{\star}(\tilde{\phi},\tilde{\psi})}(\varrho)|\le c \sqrt{\log n}
		\]
		with probability at least $1-cn^{-1}$, where $c>0$ is a constant not depending on $n$, and
		\item for some constant $c>0$ not depending on $n$, one has
		\[
		\mathbb{P}\left( 
		\sup_{(\tilde{\phi},\tilde{\psi}) \in \mathcal{F}_p}
		\frac{\max_{1\le j \le p} \hat{\sigma}_{j}^2(\tilde{\phi},\tilde{\psi})}{\sigma_{\max}^2(\tilde{\phi},\tilde{\psi})} 
		< 2 \right) \ge 1 - cn^{-1},
		\]
		where $\sigma_{\max}(\tilde{\phi},\tilde{\psi})=\max\{\sigma_{j}(\tilde{\phi},\tilde{\psi}): 1\le j\le p\}$.
	\end{enumerate}
	Consequently, if 
	\[
	\nu_0=\max_{1\leq j_1,j_2\leq \lfloor \Tn^{1/2}\rfloor}|\expect(\langle X,\phi_{j_1}\rangle_1\langle Y,\psi_{j_2}\rangle_2)|\geq c_0 \max\{\sigma_{\max}^{\tau}(\phi, \psi) n^{-1/2} \log(n), a_n\}
	\]
	for a sufficiently large constant $c_0>0$, where
	$\Tn=n^{1/(2(\bar{\alpha}+1))}$, then with $p_1\geq \lfloor h_n^{1/2}\rfloor$ and $p_2\geq \lfloor h_n^{1/2}\rfloor$, the null hypothesis in \eqref{HT-origin} is rejected uniformly over $\mathcal{F}_p$ with probability tending to one, that is,
	\[
	\mathbb{P} \left( \forall (\tilde{\phi},\tilde{\psi}) \in \mathcal{F}_p: ~
	T_{U}(\tilde{\phi},\tilde{\psi}) > q_{M^{\star}(\tilde{\phi},\tilde{\psi})}(1-\varrho/2) \text{ or } 
	T_{L}(\tilde{\phi},\tilde{\psi}) < q_{M^{\star}(\tilde{\phi},\tilde{\psi})}(\varrho/2)	
	\right) \to 1,
	\]
	as $n\to \infty$.
	\label{thm:power}
\end{theorem}

The above theorem, where the condition on $p_1$ and $p_2$ is clearly satisfied by $p_1=p_2=n$, implies that the proposed test exhibits local power of the order $a_n\vee (n^{-1/2}\log n)$. It also implies that, for any fixed alternative, the power of the proposed test converges to one as $n\rightarrow\infty$.


\section{Simulation Studies}\label{sec:simulation}

To illustrate the numerical performance of the proposed method, {we consider three families of models. For each family, we consider various settings; see below for details. In all settings, $Y$ is computed from \eqref{FLM} with $\expect Y = 1$.
	
	For each setting, we consider different sample sizes, namely, $n=50$ and $n=200$, to investigate the impact of $n$ on the power of a test. For the proposed test, we set $p_1=n$ when $d_{\mathcal X}=\infty$ and $p_2=n$ when $d_{\mathcal Y}=\infty$, i.e., we do not need to tune the parameters $p_1$ and $p_2$. The tuning parameter $\tau$ is selected by the method described in \cite{lin2020high}. Finally, we independently perform $R=1000$ replications for each setting, based on which we compute the empirical size as the proportion of rejections among the $R$ replications when the null hypothesis is true and compute the empirical power as the proportion of rejections  when the alternative hypothesis is true. In all settings, the significance level is $\varrho=0.05$.
	
	\textbf{Scalar-on-function}. The functional predictor $X$ is a centered Gaussian process with the following Mat{\'e}rn covariance function
	\[
	C(s, t) = \sigma^2 \frac{2^{1-\nu}}{\Gamma(\nu)} \left( \sqrt{2\nu} \frac{|s-t|}{\rho} \right)^{\nu} K_{\nu}\left(\sqrt{2\nu} \frac{|s-t|}{\rho} \right),
	\]
	where $\Gamma$ is the gamma function and $K_{\nu}$ is the modified Bessel function of the second kind. Here, we fix $\nu=1$, $\rho=1$ and $\sigma=1$.  The noise $Z$ is sampled from the Laplacian distribution with zero mean and unit variance, so that the distribution of $Y$ is non-Gaussian.
	\begin{figure}[t]
		\begin{centering}
			\includegraphics[scale=0.4]{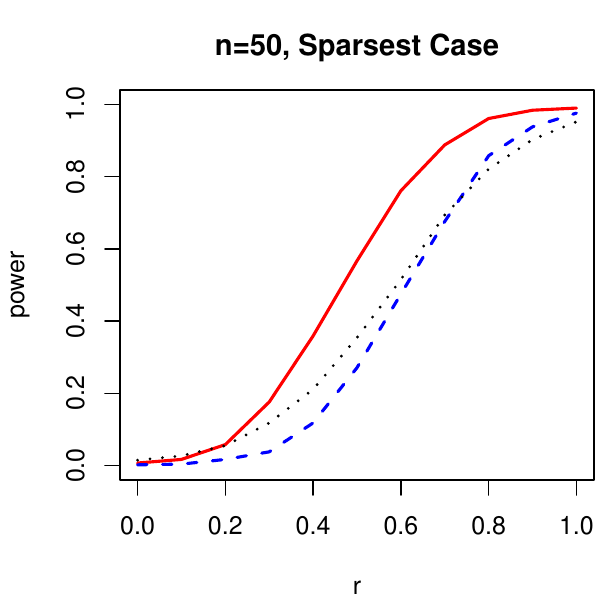}\hspace{2mm}
			\includegraphics[scale=0.4]{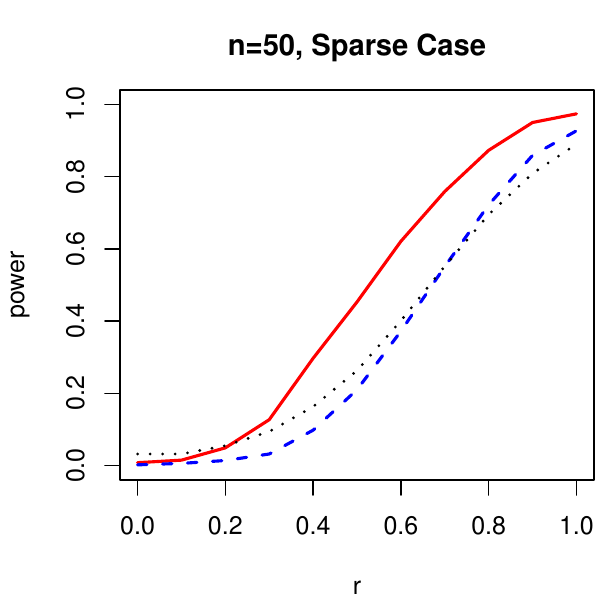}\hspace{2mm}
			\includegraphics[scale=0.4]{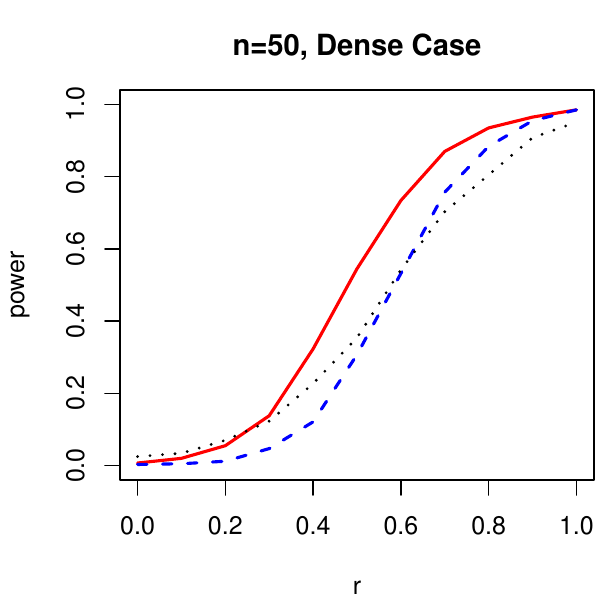}\hspace{2mm}
			\includegraphics[scale=0.4]{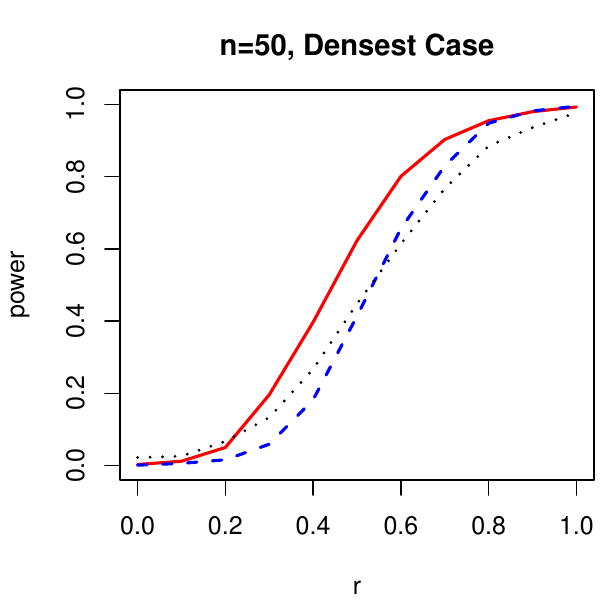}
			\par 
			\includegraphics[scale=0.4]{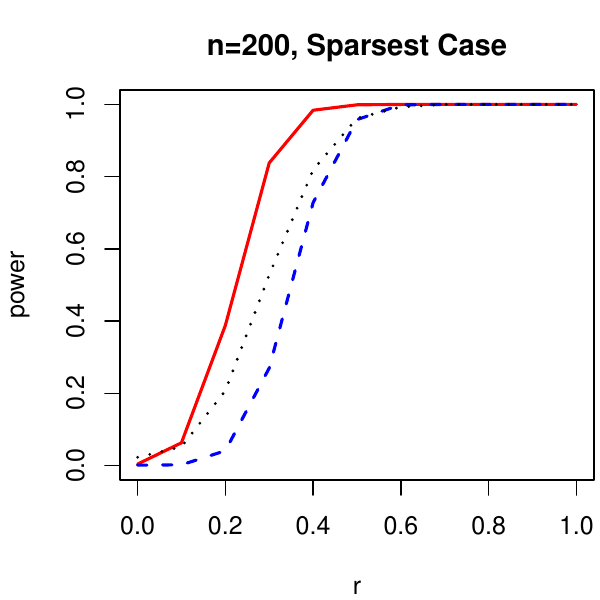}\hspace{2mm}
			\includegraphics[scale=0.4]{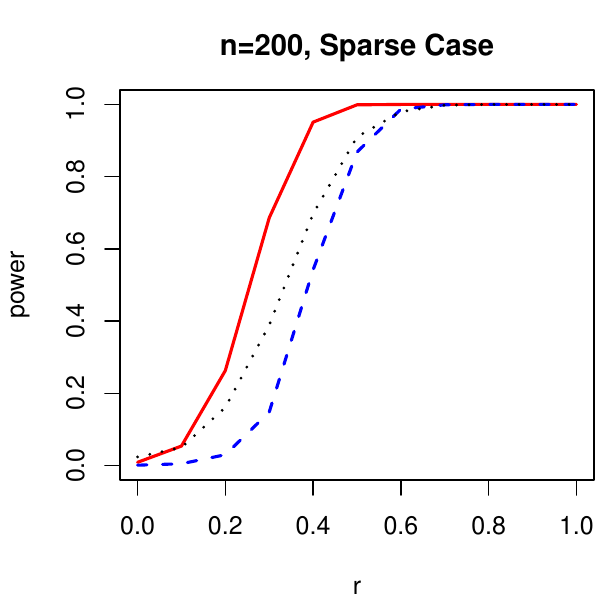}\hspace{2mm}
			\includegraphics[scale=0.4]{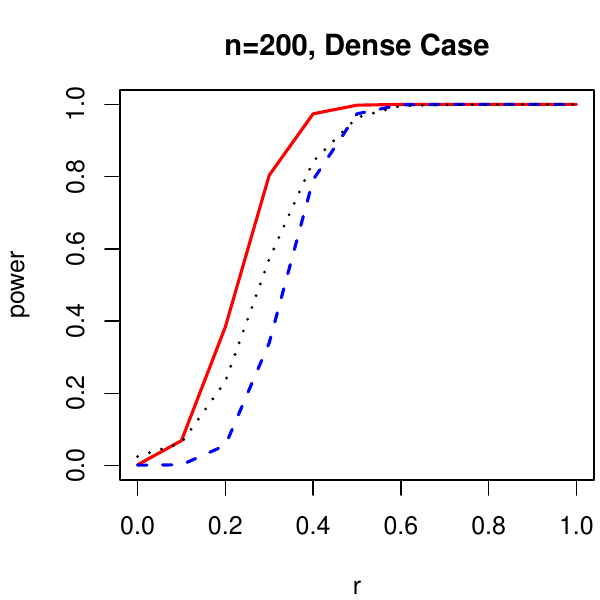}\hspace{2mm}
			\includegraphics[scale=0.4]{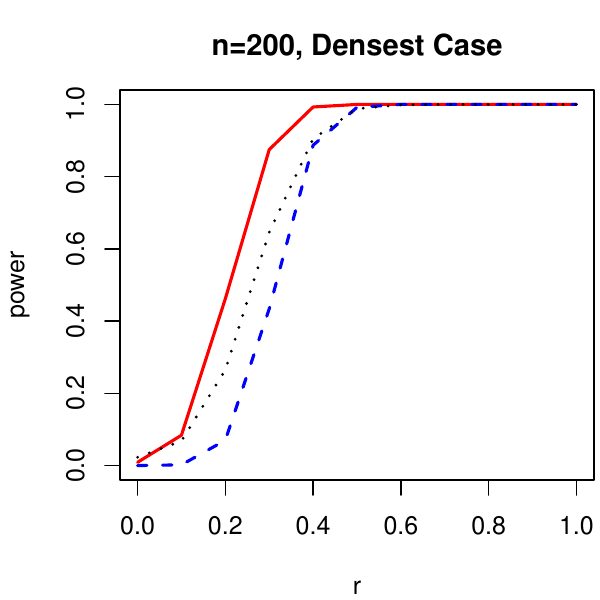}
			\par 
		\end{centering}
		\caption{Empirical size ($r=0$) and power ($r>0$) of the proposed method (red-solid), the exponential scan method (blue-dashed) and the Fisher-type method (black-dotted) for the scalar-on-function family.}
		\label{fig:sof-Laplacian}
	\end{figure}

	We consider the slope operator as $\beta(t)=r g(t)$ for $r\in\real$ with the following  distinct functions $g(t)$:
	\begin{itemize}
		\item (Sparest) $g(t)=1$;
		\item (Sparse) $g(t)=\sum_{j=1}^{3} \frac{11}{4}(j+2)^{-1}\phi_{j}(t)$; 
		\item (Dense) $g(t)=\sum_{j=1}^{K} \frac{12}{4}(j+2)^{-1}\phi_{j}(t)$ for $K=100$;
		\item (Densest) $g(t)=\frac{6}{4}t^{2}e^{t}$. 
	\end{itemize}
	The parameter $r=0,0.1,\ldots,1$ controls the strength of the signal. The case of $r=0$ corresponds to the null hypothesis, while the case of $r>0$ corresponds to the alternative hypothesis and the power of a test is expected to increase as $r$ increases. 
	In the sparse setting, $g(t)$ is formed by only a few principal components, while in the dense and densest settings, $g(t)$ contains considerably more components and thus represents challenging settings. 
	
	We compare the proposed method with the exponential scan method \citep{Lei2014} and a Fisher-type method \citep{Hilgert2013}, where for the proposed method the bases $\tilde\phi$ and $\tilde\psi$ are pragmatically taken to be the empirical eigenelements of the sample covariance operators $\hat C_X$ and $\hat C_Y$, respectively. From the results shown in Figure \ref{fig:sof-Laplacian}, we see that the proposed method  controls the empirical type-I error well and has empirical power increasing with the sample size and approaching one as the signal, quantified by $r$, becomes stronger. Moreover, the proposed test outperforms the other two methods by a large margin. 

 \begin{figure}[t]
		\begin{centering}
			\includegraphics[scale=0.4]{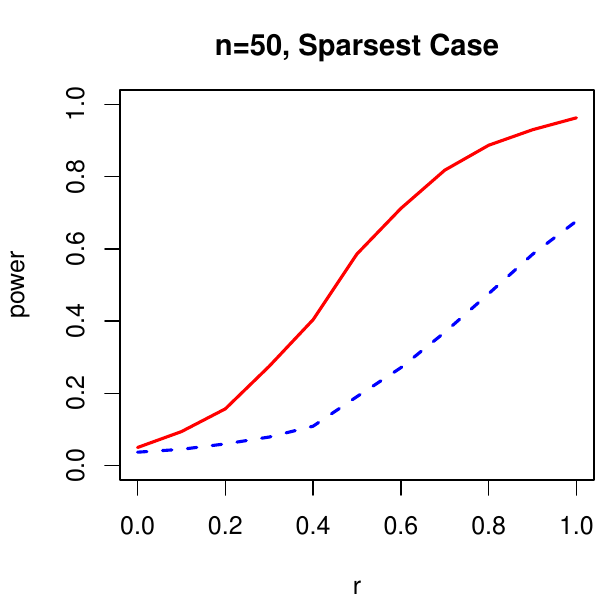}\hspace{2mm}
			\includegraphics[scale=0.4]{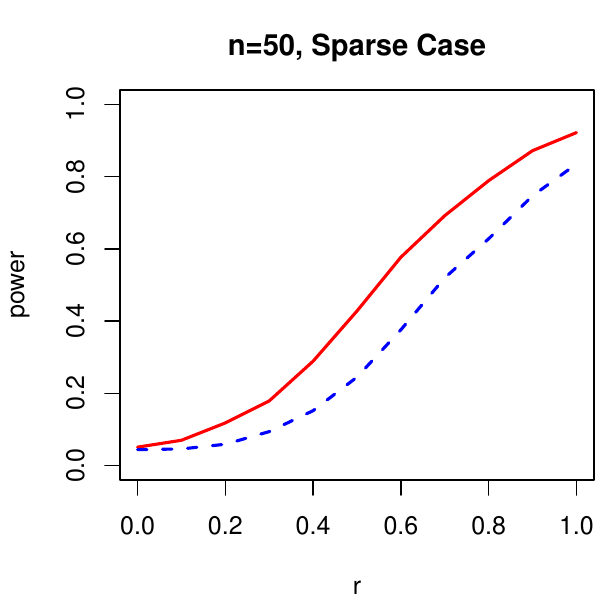}\hspace{2mm}
			\includegraphics[scale=0.4]{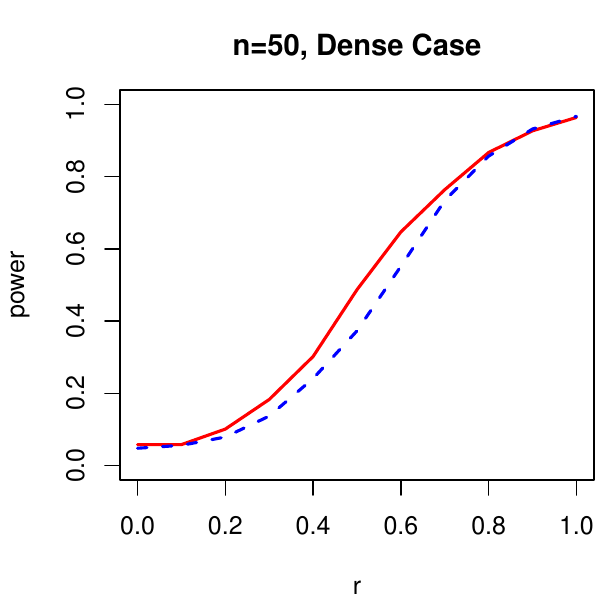}\hspace{2mm}
			\includegraphics[scale=0.4]{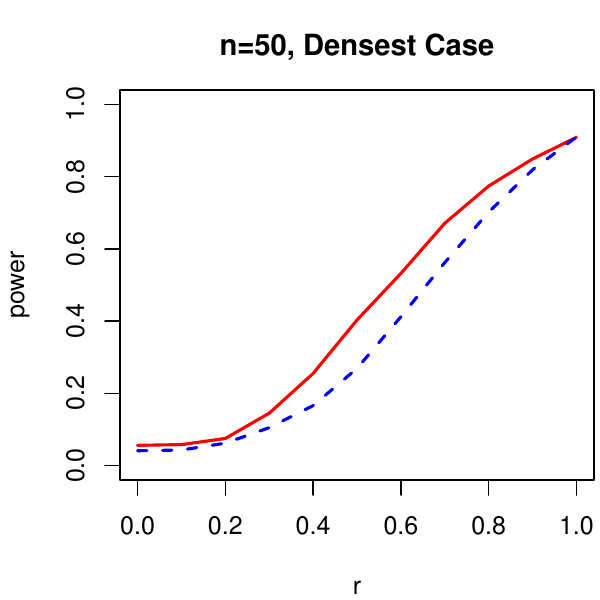}
			\par 
			\includegraphics[scale=0.4]{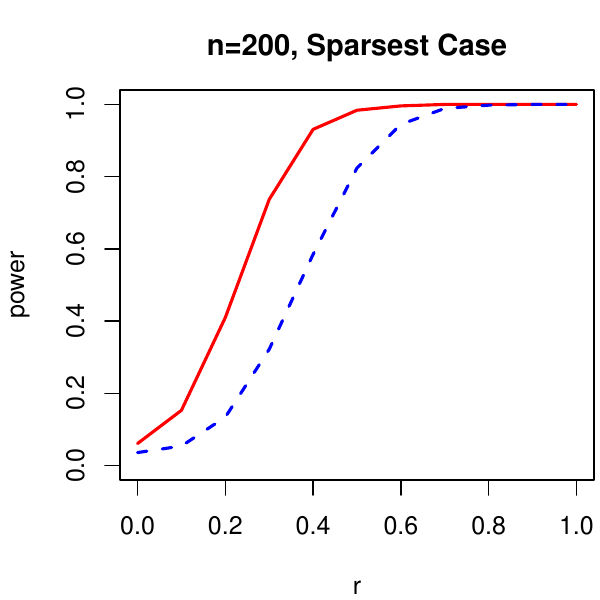}\hspace{2mm}
			\includegraphics[scale=0.4]{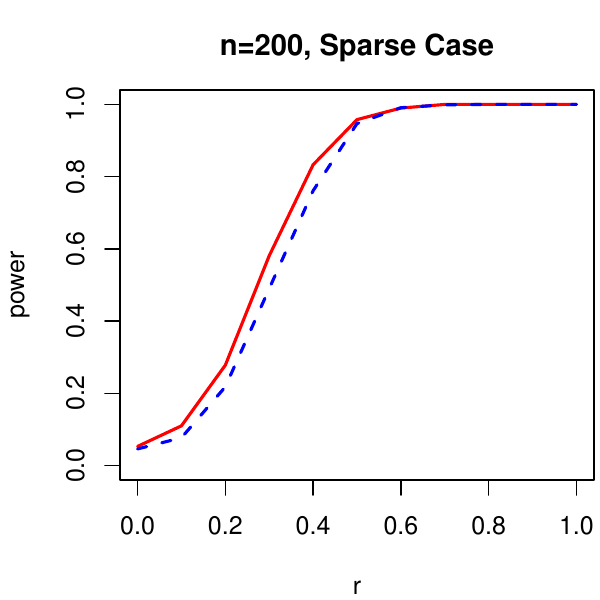}\hspace{2mm}
			\includegraphics[scale=0.4]{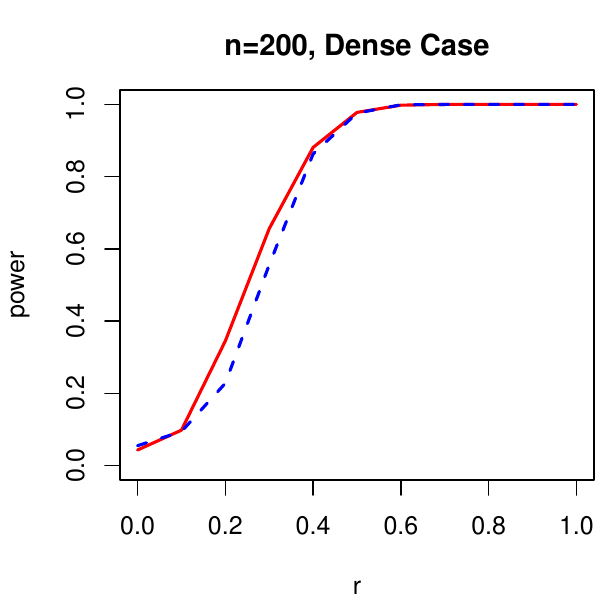}\hspace{2mm}
			\includegraphics[scale=0.4]{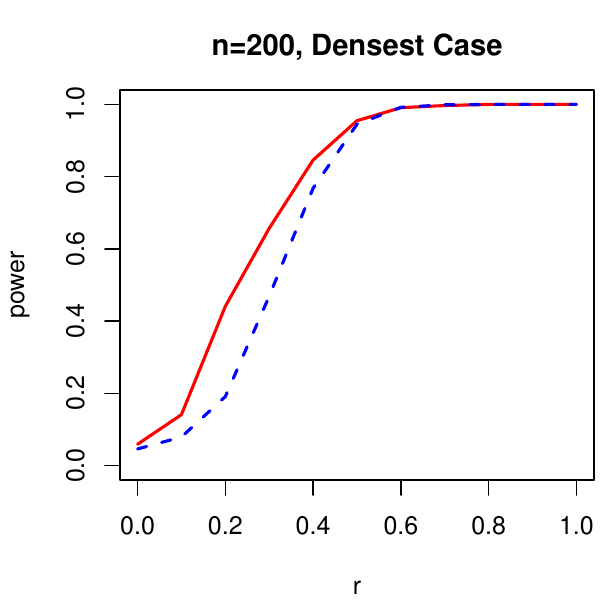}
			\par 
		\end{centering}
		\caption{Empirical size ($r=0$) and power ($r>0$) of the proposed method (red-solid) and the chi-squared test (blue-dashed) for the function-on-function family.}
		\label{fig:fof-Laplacian}
	\end{figure}

 \textbf{Function-on-function.} The functional predictor $X$ is sampled as in the scalar-on-function case, while the noise process $Z$ is represented by 
	\[
	Z(t) = \sum_{j=1}^{k} \eta^{(j)} \phi_j(t)
	\]
	for $k=50$, 
	where 
	$\phi_{1}(t)\equiv1$, $\phi_{2j}(t)=\sqrt{2}\cos(2j\pi t)$ and $\phi_{2j+1}(t)=\sqrt{2}\sin(2j\pi t)$, and for each $j=1,\ldots,k$, $\eta^{(j)}$ is a random variable following the centered Laplacian distribution  $\text{Laplace}(0, \sqrt{\lambda_j/2})$. Consequently, the process $Y$ is non-Gaussian.
	
	For the slope operator, we consider $\beta(s, t)=r g(s, t)$ with $g(s, t) \in \mathbb{R}$ being one of the following:
	\begin{itemize}
		\item (Sparsest) $g(s, t)=\frac{5}{7}$;
		\item (Sparse) $g(s, t)=\sum_{k=1}^{3}\sum_{j=1}^{3} \frac{10}{4}\frac{\phi_{j}(s)\phi_{k}(t)}{(j+2)^{1.2}(k+2)^{1.2}}$;
		\item (Dense) $g(s, t)=\sum_{k=1}^{K}\sum_{j=1}^{K} \frac{9}{4}\frac{\phi_{j}(s)\phi_{k}(t)}{(j+2)^{1.2}(k+2)^{1.2}}$ with $K=100$;
		\item (Densest) $g(s, t)=\frac{10}{4} (st)^{2}\sqrt{e^{(s+t)/2}}$.
	\end{itemize}
	The dense case represents a challenging setting as $\beta$ contains a large number of relatively weaker spectral signals, in contrast with the sparse case in which the spectral signals of $\beta$ are stronger. 
	
	We compare the proposed method with the chi-squared test \citep{kokoszka2008testing}. 
	{The chi-squared test is also based on functional principal component analysis, but unlike our method, it requires a delicate choice of the number of principal components, as the choice has a visible influence on the performance of the test. In our simulations, we take the leading $K=4$ principal components as in \cite{kokoszka2008testing}; we also tried various values for $K$ and found that overall $K=4$ yields the best results for the chi-squared test. 
	According to the results shown in Figure \ref{fig:fof-Laplacian}, the proposed method consistently outperforms the chi-squared test, and particularly has much larger power when the sample size is small or the signal is sufficiently sparse.}

 \begin{figure}[t]
	\begin{centering}
		\includegraphics[scale=0.4]{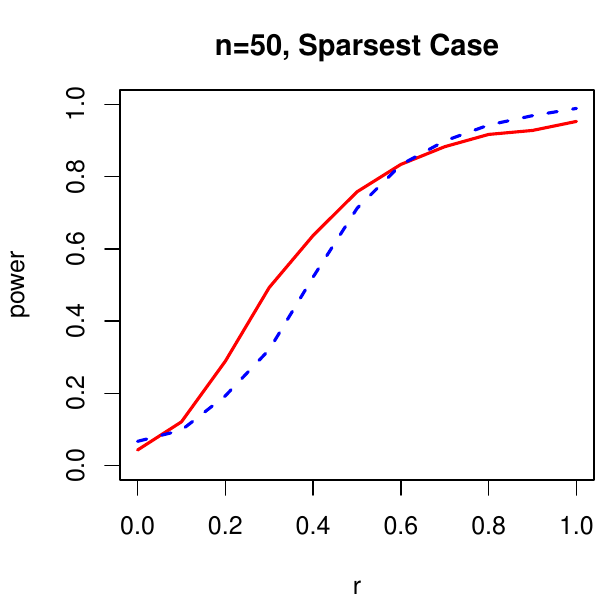}\hspace{2mm}
		\includegraphics[scale=0.4]{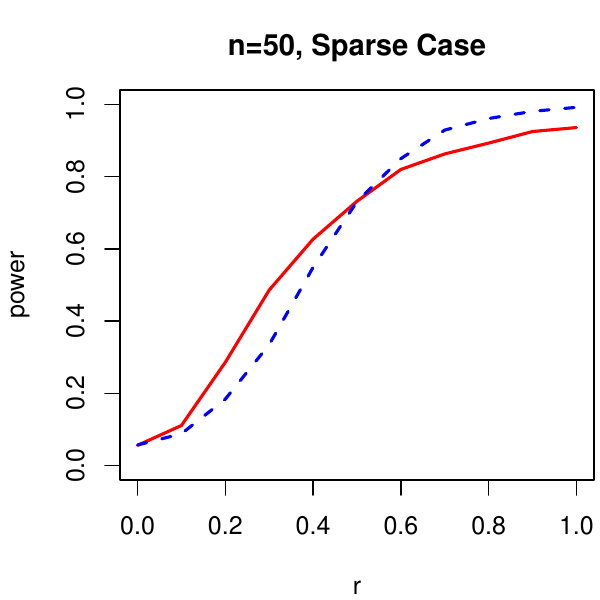}\hspace{2mm}
		\includegraphics[scale=0.4]{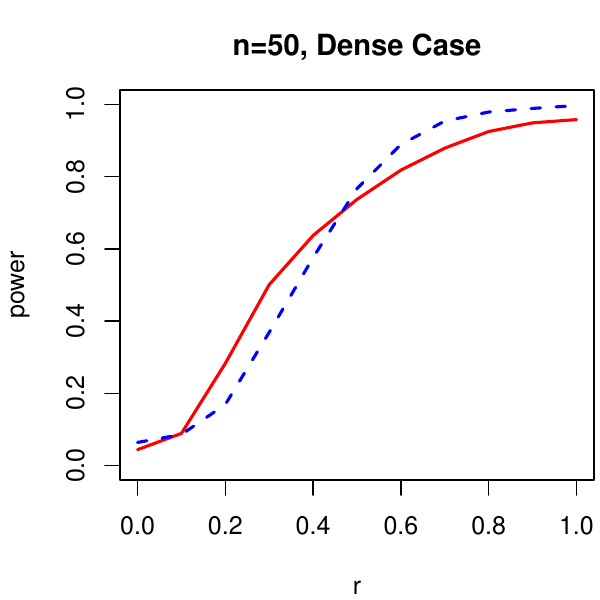}\hspace{2mm}
		\includegraphics[scale=0.4]{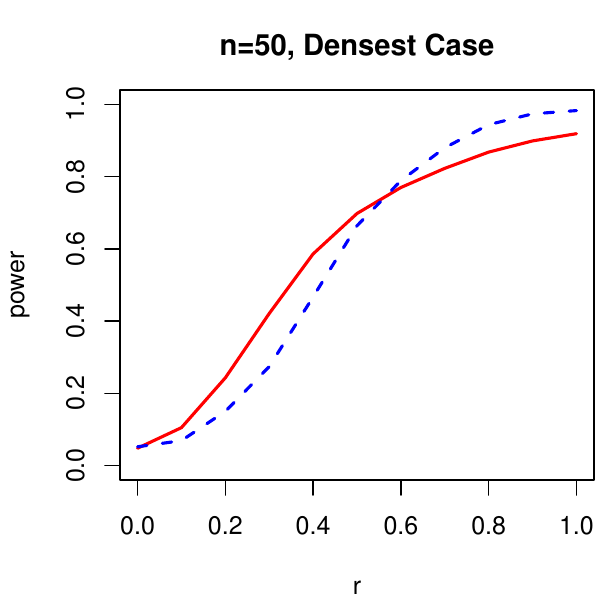}
		\par 
		\includegraphics[scale=0.4]{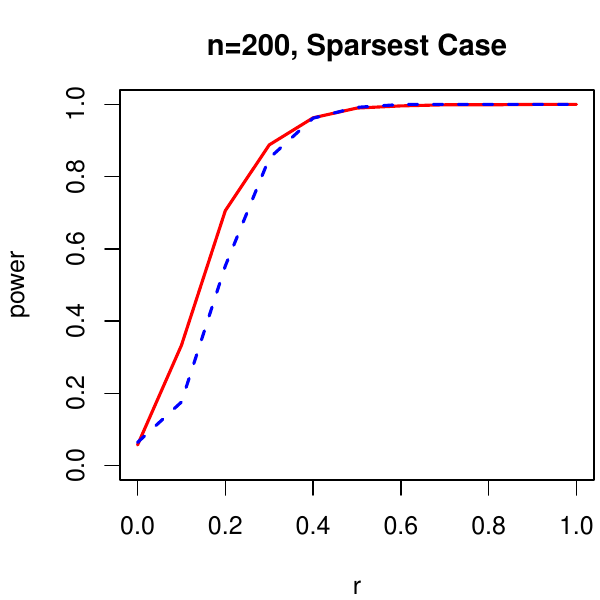}\hspace{2mm}
		\includegraphics[scale=0.4]{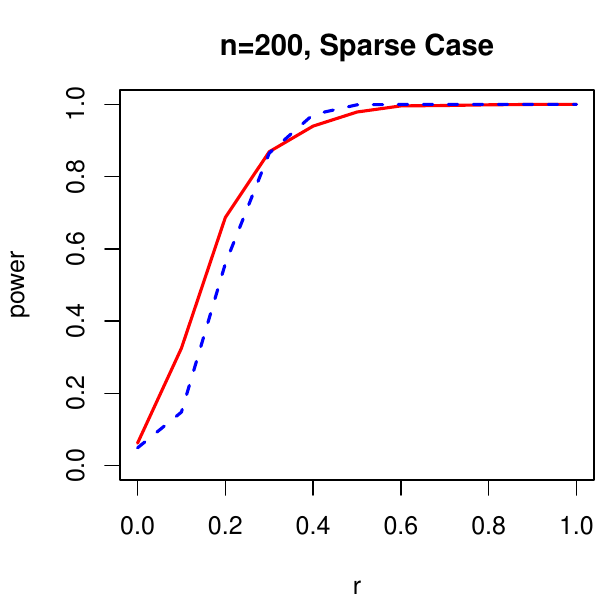}\hspace{2mm}
		\includegraphics[scale=0.4]{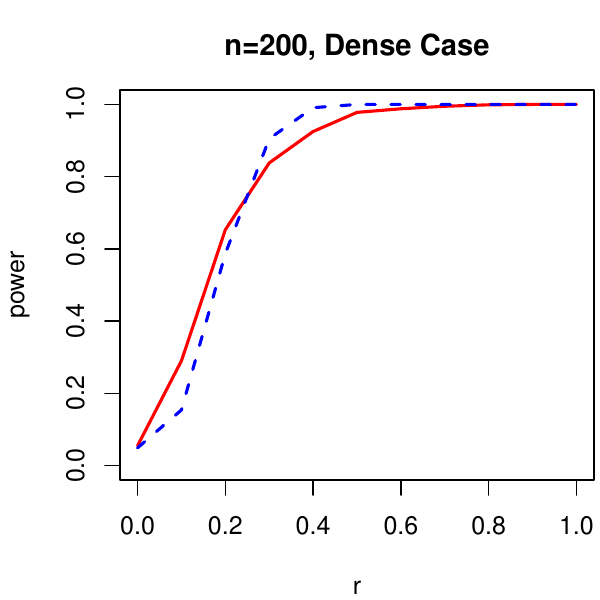}\hspace{2mm}
		\includegraphics[scale=0.4]{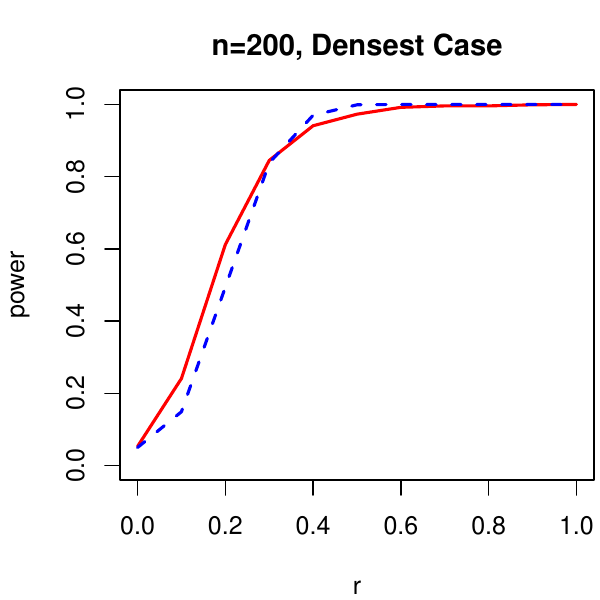}
		\par 
	\end{centering}
	\caption{Empirical size ($r=0$) and power ($r>0$) of the proposed method (red-solid) and the F-test (blue-dashed) for the function-on-vector family.}
	\label{fig:fov-Laplacian}
\end{figure}

	\textbf{Function-on-vector.} The vector predictor $X\in\real^q$ with $q=5$ follows the centered multivariate Laplacian distribution with the covariance matrix $\Sigma=A D A^{\transpose}$ for $D = \text{diag}(\lambda_{1} \ldots, \lambda_{q})$, where  $\lambda_{j}=j^{-1.5}$ and $A$ is an orthogonal matrix that is randomly generated and then remains fixed throughout the studies. The noise $Z$ is sampled as in the function-on-function case.
	
	For the slope operator, we set $\beta(t)=rg(t)$ with $g(t)\in \mathbb{R}^q$, where for each $j=1,\ldots, q$, the $j$th component $g^j(t)$ of $g(t)$ is one of the following:
	\begin{itemize}
		\item  (Sparest) $g^{j}(t)=\frac{11}{10}$;
		\item (Sparse) $g^{j}(t)=\sum_{k=1}^{3}\sum_{j=1}^{3} \frac{11}{4}\frac{\phi_{j}(\frac{j-1}{q-1})\phi_{k}(t)}{(j+2)^{1.2}(k+2)^{1.2}}$;
		\item (Dense) $g^{j}(t)=\sum_{k=1}^{K}\sum_{j=1}^{K} \frac{6}{4}\frac{\phi_{j}(\frac{j-1}{q-1})\phi_{k}(t)}{(j+2)^{1.2}(k+2)^{1.2}}$ with $K=100$;
		\item (Densest) $g^{j}(t)= \frac{11}{4} (\frac{j-1}{q-1})^{2}\sqrt{e^{\frac{j-1}{q-1}/4}} t^{2}\sqrt{e^{t/4}}$.
	\end{itemize}
	Similar to the function-on-function family, the dense case represents a more challenging setting for the proposed method. 
	We compare the proposed method with the F-test developed by \cite{Zhang2011}. From the results shown in Figure \ref{fig:fov-Laplacian}, we observe that the proposed method is more powerful when the signal is relatively weak while the F-test has slightly higher power when the signal is strong.}

\section{Data Application}\label{sec:applications}

We apply the proposed method to study physical activities using data collected from wearable devices and available in the National Health and Nutrition Examination Survey (NHANES) 2005-2006. Over seven consecutive days, each participant wore a wearable device that for each minute recorded the average physical activity intensity level (ranging from 0 to 32767) in that minute. As the wearable devices were not waterproof, participants were advised to remove the devices when they swam or bathed. The devices were also removed when the participants were sleeping. For each subject, an activity trajectory, denoted by $A(t)$ for $t\in[0,7]$, was collected. In our study, trajectories with missing values or unreliable readings are excluded.

To eliminate the effect of circadian rhythms that vary among participants, instead of the raw activity trajectories, we follow the practice in \cite{Chang2020+, lin2020high} to consider the activity profile $Y(s)=\mathrm{Leb}(\{t\in[0,7]:A(t)\geq s\})$ for $s=1,\ldots,32767$, where $\mathrm{Leb}$ denotes the Lebesgue measure on $\real$. The zero intensity values are also excluded since they may represent no activities like sleeping or intense activities like swimming. After these pre-processing steps, for the $i$th subject, we obtain an activity profile $Y_i(s)$ which is regarded as a densely observed function. Our goal is to study the effect of age on the activity profile. As children and adults, as well as males and females, have different activity patterns, we conduct the study on each group separately by using the proposed test, where the tuning parameter $\tau$ is selected by using the method of \cite{lin2020high}.

First, we consider children with age from 6 to 17, including 6 and 17, and focus on the intensity spectrum $[1,1000]$ as  children are found to have more moderate activities \citep{WHO2020}. As shown in Table \ref{tab:real-data-VC-pvalue}, the age seems no impact on the activity profile for female children, but has significant impact for male children. By inspecting the mean activity profile curves in Figure \ref{fig:real-data-VC-children}, we see the visible differences for different age groups among male children, in contrast with the visually indistinguishable differences among female children. In particular, our test results and Figure \ref{fig:real-data-VC-children} together suggest that on average young male children tend to be significantly more active than elder male children.

\begin{table}[t]
	\centering
	\caption{The p-values for testing the effect of age on the activity pattern, with sample size in the parentheses.}
	\label{tab:real-data-VC-pvalue}
	\begin{tabular}{|c|c|c|}
		\hline
		& Male & Female \\
		\hline
		p-value (age 6-17)  & 0.0046 (962) & 0.1778 (952) \\
		\hline
		p-value (age 18-35) & 0.1336 (623)  & 0.0282 (823) \\
		\hline
	\end{tabular}
\end{table}

\begin{figure}[t]
	\begin{center}
		\includegraphics[scale=0.5]{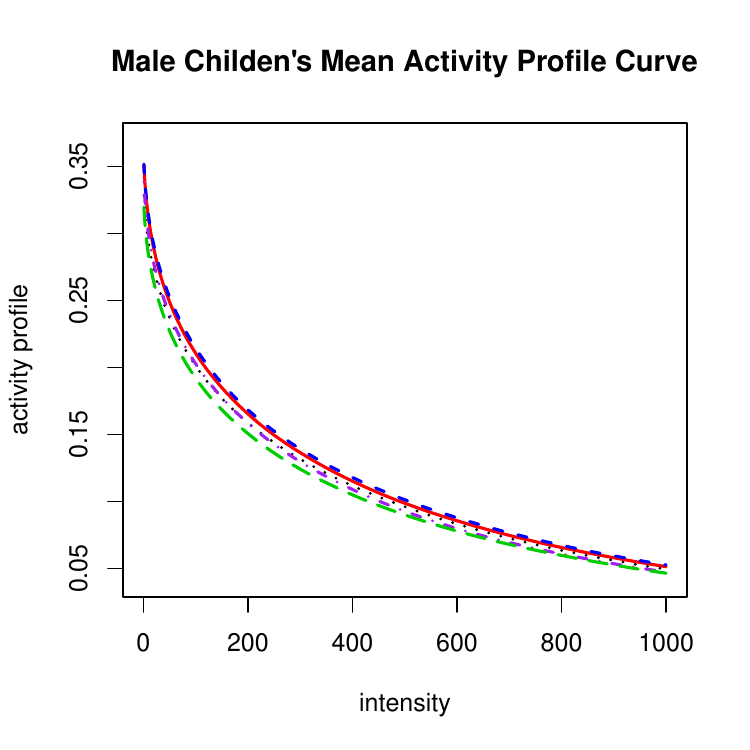}\quad{}
		\includegraphics[scale=0.5]{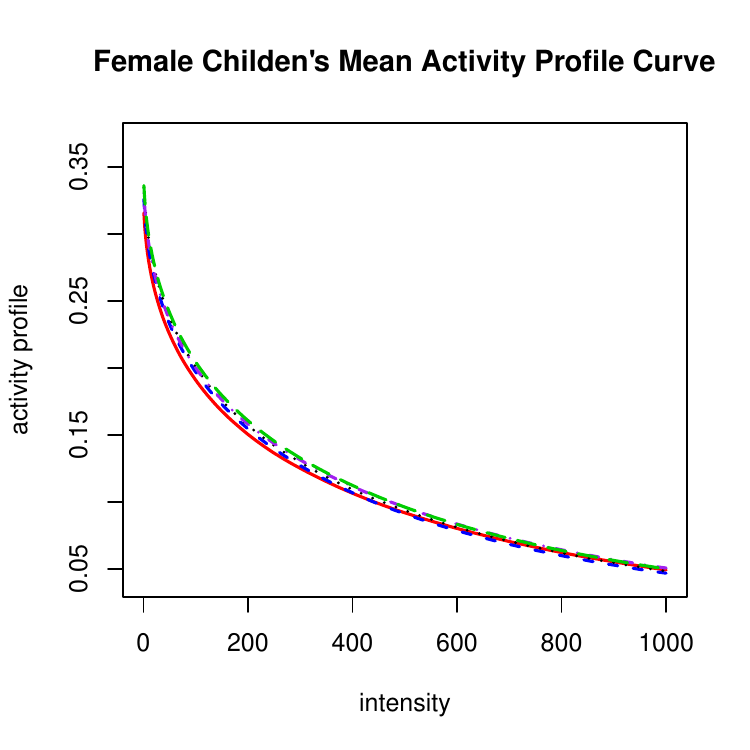}
	\end{center}
	\caption{Mean activity profile curves among male children (left) and female children (right) for  different age groups, namely, age 6-8 (red-solid), age 9-10 (blue-dashed), age 11-12 (black-dotted), 13-14 (purple-dash-dotted) and age 15-17 (green-dashed).}
	\label{fig:real-data-VC-children}
\end{figure}

Now we consider the young adults with age from {18 to 35}, and focus on the intensity spectrum $[1,3000]$. As shown in Table \ref{tab:real-data-VC-pvalue}, there is significant difference of mean activity profiles among  female young adults, while the difference is not significant among the males. This also agrees with the mean activity profiles shown in Figure \ref{fig:real-data-VC-young-gender}, where we observe significant difference in the mean activity profiles among different age groups of females, especially on the intensity spectrum $[300,1500]$, while the difference is less pronounced among males.

\begin{figure}[t]
	\begin{center}
		\includegraphics[scale=0.5]{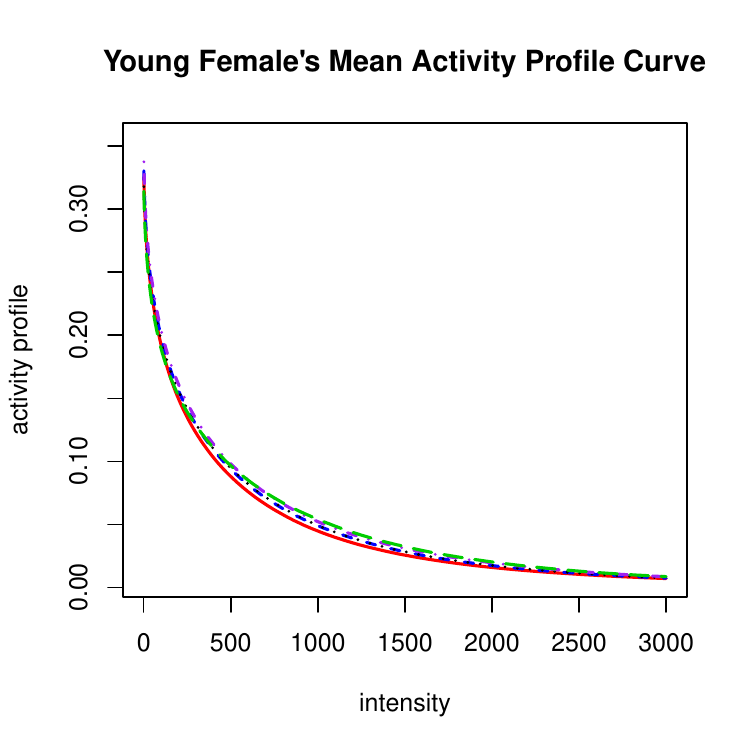}\quad{}
		\includegraphics[scale=0.5]{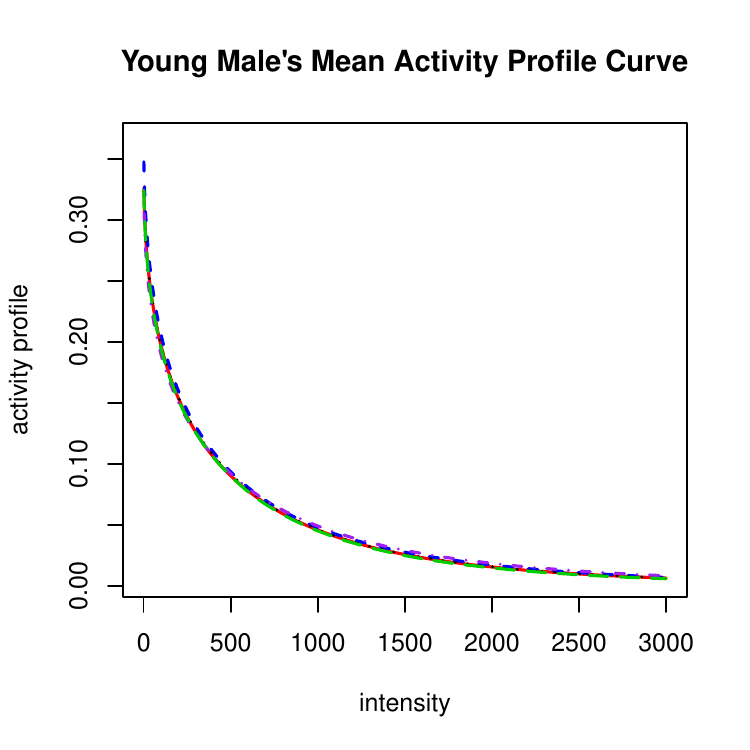}
		\par 
		\includegraphics[scale=0.5]{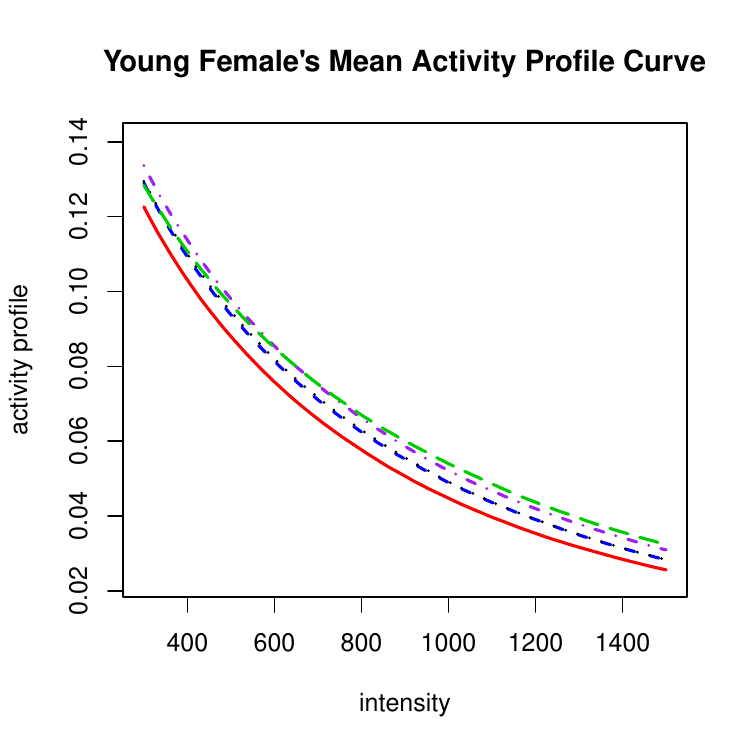}\quad{}
		\includegraphics[scale=0.5]{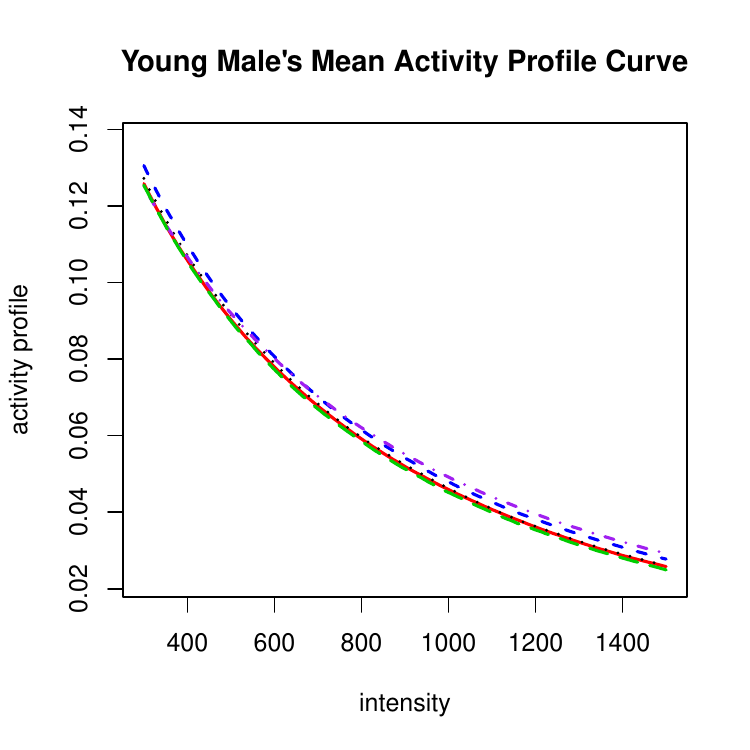}
	\end{center}
	\caption{Mean activity profile curves among young female (top-left) and male (top-right) adults with their zoom-in regions (bottom) on the intensity spectrum $[300,1500]$ in different age groups, namely, age 18-21 (red-solid), age 22-25 (blue-dashed), age 26-29 (black-dotted),  age 30-33 (purple-dash-dotted) and age 33-35 (green-dashed).}
	\label{fig:real-data-VC-young-gender}
\end{figure}

\section{Concluding Remarks}\label{sec:dis}

In this paper, we developed a novel approach for testing nullity of the slope function in functional linear regression. This method fully circumvents the challenge of ill-posedness without requiring an intricate choice of the number of principal components, thereby enhancing numeric stability and potentially improving the test's power. 
We also uncovered an interesting distinction between estimating and inferring the slope function. Specifically, for estimation to be consistent, it can incorporate no more than $\sqrt n$ empirical principal components. However, for the inference purpose, the method allows the use of all empirical principal components while maintaining its  asymptotic validity and consistency. To the best of our knowledge, our test is the first of its kind to utilize all $n$ empirical principal components.

While we focused on fully or densely observed functional data in this work, in real-world applications, there are applications with sparsely observed data, such as longitudinal/panel data in medicine and econometrics. Adapting our framework to accommodate such sparse data presents a challenging yet worthwhile future research direction. 
Moreover, investigating the theoretical transition between densely and sparsely observed data is intriguing. This exploration could yield valuable insights for practical data applications.

\section*{Acknowledgement}
This research is partially supported by NUS startup grant A-0004816-01-00 and MOE AcRF Tier 1 grant  A-0008522-00-00.

\section*{Supplementary Material}\label{SM}

Supplementary material contains proofs for the theoretical results of this paper, some auxiliary results used in the proofs and a concentration inequality for empirical eigenelements. (PDF)

\bibliographystyle{asa}
\bibliography{ref}

\includepdf[page=-]{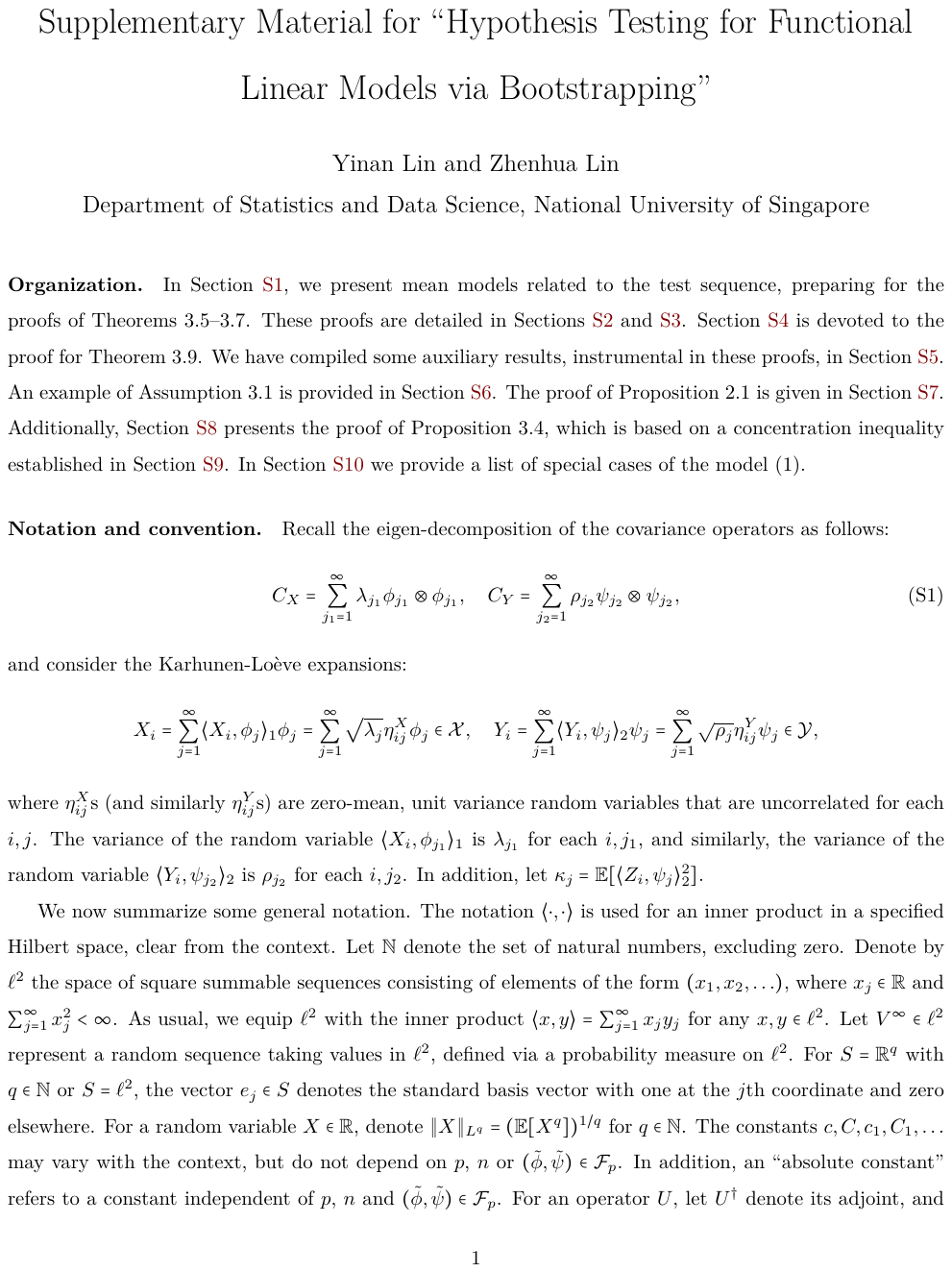}

\end{document}